\documentclass[fleqn,usenatbib]{mnras}

\usepackage[T1]{fontenc}

\DeclareRobustCommand{\VAN}[3]{#2}
\let\VANthebibliography\thebibliography
\def\thebibliography{\DeclareRobustCommand{\VAN}[3]{##3}\VANthebibliography}

\usepackage{graphicx}	
\usepackage{amsmath}	
\usepackage{amssymb}	
\usepackage{dblfloatfix}
\usepackage{caption}
\usepackage{nccmath}
\usepackage{comment}
\usepackage{multirow}
\usepackage{newtxtext,newtxmath}

\usepackage[justification=centering]{caption}

\defcitealias{sellwood22}{SS22}
\defcitealias{mancera24}{MP24}

\title[]{The long life of ultra diffuse galaxies inside low-density dark matter halos: the case of AGC 114905}

\author[A. Afruni et al.]{
Andrea Afruni,$^{1,2}$\thanks{E-mail: andrea.afruni@unifi.it}
Federico Marinacci,$^{3,4}$\
Pavel E. Mancera Pi\~na,$^{5}$\ \&
Filippo Fraternali$^{1}$
\\
$^{1}$Kapteyn Astronomical Institute, University of Groningen,Landleven 12, 9747 AD Groningen, The Netherlands\\
$^{2}$Dipartimento di Fisica e Astronomia, Universit\`a di Firenze, Via G. Sansone 1, 50019 Sesto Fiorentino, Firenze, Italy\\
$^{3}$Department of Physics and Astronomy, University of Bologna, Via P. Gobetti 93/2, I-40129 Bologna, Italy\\
$^{4}$INAF, Astrophysics and Space Science Observatory Bologna, Via P. Gobetti 93/3, I-40129 Bologna, Italy\\
$^{5}$Leiden Observatory, Leiden University, P.O. Box 9513, 2300 RA, Leiden, The Netherlands
\\
}

\date{Accepted XXX. Received YYY; in original form ZZZ}

\pubyear{2025}

\begin{document}
\label{firstpage}
\pagerange{\pageref{firstpage}--\pageref{lastpage}}
\maketitle

\begin{abstract}
It has long been known that, in the absence of a dark matter (DM) halo, galaxy discs tend to develop global gravitational instabilities that strongly modify their initial structure. The recent discovery of gas-rich ultra diffuse galaxies (UDGs) that seem to live in DM halos with very low concentrations, a very atypical configuration in the standard cosmological framework, poses therefore a crucial question: is the small contribution from such DM halos sufficient to stabilize the UDG discs? In this work we investigate this question, focusing on the extreme UDG $\rm{AGC\ 114905}$, which previous works found to be unstable. Here, we revisit these studies, using idealised numerical simulations with AREPO of a system composed by a stellar disc, a gas disc and a DM halo in initial equilibrium with each other and with properties based on slightly revised observational data of $\rm{AGC\ 114905}$. We explore different scenarios for the DM halo and we run our simulations for 5 Gyr. We find that in all cases the stellar and the gas discs are stable and that their initial density distributions and kinematic properties remain unchanged during the course of the simulation. We discuss how the apparent discrepancy with previous works (where the UDG developed instabilities) is due to our discs being dynamically hotter and living in slightly more massive DM halos, in accordance with the new observational constraints, previously unavailable. Our findings demonstrate that $\rm{AGC\ 114905}$ (and likely other similar UDGs) can evolve unperturbed in halos that challenge current cosmological models.
\end{abstract}




\section{Introduction}\label{intro}

The persistence of a galactic disc against global gravitational instabilities that could dramatically modify its initial structure (generally leading to the formation of a prominent central bar), has been investigated for more than five decades, both through linear perturbation analyses \citep[e.g.][]{toomre81,efstathiou82} and with numerical simulations \citep[e.g.][]{hohl71,sellwood16,teppergarcia21}. While the subject is extremely complex \citep[see e.g.][]{romeo23}, a robust result of these stability analyses is that an additional stabilizing mass component, like a dark matter (DM) halo, is generally needed to keep the galactic disc from developing global instabilities. This has actually been one of the first and most striking indications of the presence of dark matter in the Universe \citep{ostriker73}.

It is therefore intriguing to study whether galaxies that appear to have very low dark matter contents can actually be stable against global instabilities. One remarkable science-case is related to ultra diffuse galaxies, or UDGs. These objects were originally defined by \citeauthor{vandokkum15} (\citeyear{vandokkum15}, but see also \citealt{impey97,conselice18,chamba20}) as low surface brightness galaxies ($\mu_{0,g}\gtrsim24$ mag$/\rm{arcsec}^2$) with unusually large sizes (effective radii larger than about $1.5$ kpc) and due to their unique properties they have been an intensive subject of study in the last decade both from an observational \citep[e.g.][]{roman17,mancera18,vandokkum18} and a theoretical \citep[e.g.][]{dicintio17, chan18,benavides23} point of view. Of particular interest for this work is a subclass of UDGs which appear to host large amounts of neutral hydrogen \citep[HI, e.g.][]{leisman17,janowiecki19}. This HI-rich UDGs generally live in isolated environments and both unresolved \citep[e.g.][]{spekkens18,hu23} and resolved HI data \citep[e.g.][]{mancera19} have shown that they appear to be rotating with velocities lower than what one would expect for objects of this size. To explain such peculiar kinematics, rotation curve decomposition analyses have been carried out showing that these galaxies reside in DM halos with unexpectedly low concentrations and therefore with very low inner densities \citep[e.g.][]{mancera19,sengupta19,shi21,kong22}. Despite the effort put in the last years into analyzing and interpreting these data, UDG observations remain challenging and the parameters of the recovered DM halos are necessarily uncertain. Studying whether both the stellar and the gas discs of these UDGs can survive in such low-density halos against global gravitational instabilities is therefore necessary and could represent an independent way to confirm the observational results listed above. This is the main motivation for the present study.

The most extreme example of these HI-rich UDGs is $\rm{AGC\ 114905}$, which was originally analysed in detail in \cite{mancera22}. These authors found that, within the uncertainties, the galaxy's rotation curve could be explained just by taking into account the gravitational potential of its baryonic components. Hence, in the context of cold dark matter (CDM), only an extremely atypical halo could be fitted \citep[but see also][]{lelli24}, with masses that would imply a much higher baryon fraction (at least of the order of the cosmological average, $f_{\rm{bar}}\approx0.16$, see \citealt{planck20}) than in typical dwarfs \citep[$\sim0.1-0.2f_{\rm{bar}}$, e.g.][]{read17,mancera22flaring} and concentrations that would be more than $10\sigma$ below the expectations of the concentration-mass relation \citep[][]{dutton14}, extreme values that are unheard of in CDM cosmological simulations \citep[e.g.][]{kong22,nadler23}.

The stability of this extreme galaxy has recently been investigated by \citeauthor{sellwood22} (\citeyear{sellwood22}, hereafter \citetalias{sellwood22}). They performed N-body numerical simulations of a system composed of two collisionless discs (one representing the stellar and one the gas component), with initial properties resembling the observations of \cite{mancera22}. They explored both a case without DM halo and a case with an atypical DM halo with a low concentration (the one proposed by \citealt{mancera22} for $\rm{AGC\ 114905}$). In both cases, they found that the two discs of stars and gas get quickly (within a Gyr) disrupted by global gravitational instabilities, dramatically modifying the initial surface densities and kinematics of the UDG. Given that the HI observations of $\rm{AGC\ 114905}$ show a rather regular velocity pattern typical of a stable disc galaxy, the main conclusion of \citetalias{sellwood22} was that this UDG needs a more significant DM halo that would stabilize it. They proposed as likely explanation that the inclination of the UDG had been overestimated by \cite{mancera22}: a lower inclination would indeed imply larger rotational velocities and therefore the need for a stronger contribution from the DM halo. However, \citetalias{sellwood22} also noted that a slightly more massive halo than what they employed in their fiducial simulations, combined with a larger velocity dispersion of the two disky components, could bring the galaxy much closer to stability.

Very recently, \citeauthor{mancera24} (\citeyear{mancera24}, hereafter \citetalias{mancera24}), have revisited the findings of \cite{mancera22} on $\rm{AGC\ 114905}$. Thanks to new and unprecedentedly deep optical data these authors have independently confirmed the estimate of the UDG's inclination (which is crucial to eventually determine the galaxy's DM halo), which was previously based only on the HI data. Moreover, an updated kinematic modelling (see Section~\ref{data} for more details) led \citetalias{mancera24} to find slightly larger values of the mass of the DM halo (although still with extremely low concentrations) and of the HI velocity dispersion. These findings therefore imply that a revision of the stability analysis of \citetalias{sellwood22}, and of their main conclusions outlined above, is needed. A linear stability analysis \citep{bacchini24} has already shown that, considering the updated observational constraints of \citetalias{mancera24}, $\rm{AGC\ 114905}$ appears to be stable against local gravitational instabilities, but a full new numerical investigation of the global stability of this UDG, in light of the new observational data, is still missing. Such investigation represents the main goal of the present study. We stress here that proving the global stability of this galaxy would show that the observational data are dynamically plausible, further implying the possible existence in the Universe of DM halos that can currently hardly be explained in the usual CDM framework and can hence be used to provide clues on the nature of dark matter itself \citep[e.g.][]{nadler23}.

This paper is organized as follows: in Section~\ref{data} we summarize the main characteristics of the observational data on which we base our simulations and the results of the rotation curve decomposition of \citetalias{mancera24}; in Section~\ref{methods} we describe the procedure used to build the initial conditions of our numerical experiments and the main features of the code used for our analysis; in Section~\ref{results} we show the main results of our fiducial simulations of the UDG $\rm{AGC\ 114905}$; in Section~\ref{discussion} we compare our findings with previous works, focusing in particular on \citetalias{sellwood22} and on the results of previous linear local and global stability analyses, and we discuss the implications of our results regarding the formation of UDGs and alternative dark matter theories; finally, in Section~\ref{conclusions} we summarize the main results of this work and we outline our conclusions.

\section{Overview of the data}\label{data}
The observational constraints that we use in this study are based on the work of \citetalias{mancera24}, who studied the stellar optical emission of AGC 114905 by analysing ultra-deep images in the $g$, $r$ and $i$ bands, obtained using the Optical System for Imaging and low-Intermediate-Resolution Integrated Spectroscopy \citep[OSIRIS,][]{cepa2000} imager of the 10.4–m Gran Telescopio Canarias (GTC). The HI data were instead originally obtained by \cite{mancera22} and consist of a data cube given by the combination of observations with the Karl G. Jansky Very Large Array (VLA) using the D-, C-, and B-array configurations. We refer to \citetalias{mancera24} and to \cite{mancera22} for additional details on the data acquisition and reduction of both the optical and HI data, while here we summarize the main properties of AGC 114905 resulting from their analysis.

Through both isophotal fitting and a custom version of the modified Hausdorff distance method \citep[e.g.][]{montes19}, \citetalias{mancera24} first estimated the galaxy position angle, P.A. $=(78\pm5)^{\circ}$, and inclination, $i=(31\pm2)^{\circ}$, confirming the previous estimates obtained with HI data \citep{mancera22}. They then extracted the surface brightness and color profiles from the images in the three bands. Finally, by converting the surface brightness profile into physical units \citep[see][]{cimatti19} and adopting the mass-to-light ratio relations of \cite{du20}, they obtained the stellar mass surface density profile $\Sigma_{\rm{stars}}(R)$, which we show in the top panel of Figure~\ref{fig:ICagama} (orange data-points).

The distance $D=78.7\pm1.5$ Mpc was estimated from the galaxy systemic velocity extracted from the global profile and the kinematic modelling (see below) of the HI data. The observed HI surface density profile $\Sigma_{\rm{gas}}$ is shown in the first panel of Figure~\ref{fig:ICagama} (blue points). The kinematic modelling of the HI data was performed using the software $^{3\rm{D}}$Barolo \citep{diteodoro15}, which allows to fit the HI datacube directly in 3D and it is therefore not affected by beam smearing effects. In the present work, we use the results of \citetalias{mancera24}, who improved the kinematic modelling of \cite{mancera22}, by considering the updated position angle and inclination obtained with the analysis of the new optical images (see above) and including also faint HI emission that was previously masked. The best-fit circular velocity $v_{\rm{circ}}$ and velocity dispersion $\sigma$ profiles are shown respectively in the second and third panels of Figure~\ref{fig:ICagama} (gray and blue data points), with the values of the circular velocity obtained from the best-fit rotational velocity $v_{\rm{rot}}$ through:
\begin{ceqn}
\begin{equation}\label{eq:adrift}
v^2_{\rm{circ}} = v^2_{\rm{rot}} + \frac{R}{\rho}\frac{\partial(\rho \sigma^2)}{\partial R}\ ,
\end{equation}
\end{ceqn}
where $R$ is the cylindrical radius, $\rho$ is the density of either the gas or the stellar component and the second term is the square of the so-called asymmetric drift \citep[e.g.][]{binney08,Iorio17,mancera24}, which corrects the rotational velocities for pressure support.

We finally note that, with respect to the previous modelling of \cite{mancera22}, the updated $v_{\rm{rot}}$ and $\sigma$ are respectively $10\%$ and $40\%$ higher on average (with $\sigma$ being significantly higher especially for $R<5$ kpc), with possibly important consequences regarding the galaxy stability, as we will see below (Section~\ref{results}). 

\subsection{Decomposition of the rotation curve}\label{rotcurvedec}
\citetalias{mancera24} utilized the observed HI kinematics to constrain the properties of the DM halo of $\rm{AGC\ 114905}$, under the assumption that the circular velocity $v_{\rm{circ}}$ is a direct tracer of the total gravitational potential of the system \citep[e.g.][]{binney08}.

In order to do this, they first fitted the stellar and the gas surface densities respectively with a poly-exponential disc of fourth degree and with a function that describes an exponential disc with a depression in the inner regions \citep[see also][]{oosterloo07}. Once the surface densities were determined, they computed numerically the corresponding potentials, to infer the contribution of the baryonic components to the total gravitational potential. They then assumed a functional profile for the DM halo, in particular a \textsc{core}NFW profile \citep{read16a,read16b}:

\begin{ceqn}
\begin{equation}\label{eq:coreNFW}
\rho_{\textsc{core}\rm{NFW}}(r) = f^n \rho_{\rm{NFW}}(r)+\frac{nf^{n-1}(1-f^2)}{4\pi r^2 r_{\rm{c}}}M_{\rm{NFW}}(r)\ ,
\end{equation}
\end{ceqn} 
where $n$ is fixed to be equal to 1, $f=\tanh{(r/r_{\rm{c}})}$ is a function that creates a core of size $r_{\rm{c}}$ in the usual NFW profile, described by a density \citep{nfw96}
\begin{ceqn}
\begin{equation}\label{eq:NFW}
\rho_{\rm{NFW}}(r) = \frac{4\rho_{\rm{s}}}{(r/r_{\rm{s}})(1+r/r_{\rm{s}})^2}\ ,
\end{equation}
\end{ceqn}
and a mass profile $M_{\rm{NFW}}(r)$. In equation~\eqref{eq:NFW}, $\rho_{\rm{s}}= \rho_{\rm{NFW}}(r_{\rm{s}})$, with $r_{\rm{s}}$ being the scale radius, also used to define the halo concentration, $c_{200} = r_{200}/r_{\rm{s}}$, where $r_{200}$ is the radius inside which the average density is 200 times the critical density of the Universe. By fitting the circular velocity with such a profile and assuming a CDM framework, \citetalias{mancera24} found two possible best-fit models, which they denote as Case 1 and Case 2 (see their Figure 9, and their Table 1). The main difference between the two cases is a more stringent prior in the halo mass for Case 2, such that the baryon fraction can not be higher than the cosmological value. As a result, in Case 1 the DM halo has a lower mass ($\log(M_{200}/M_{\odot})=9.36$), so that the baryon fraction is about 4 times larger than the cosmological value, while in Case 2 (where the cosmological baryon fraction is set as an upper limit), the halo mass is slightly larger ($\log(M_{200}/M_{\odot})=10.06$). As already mentioned, the main result in both cases is that the resulting concentration ($c_{200}=2.76$ and $1.2$, respectively) is too low compared to the expectations from CDM models \citep{dutton14} for halos of this mass. This implies that this UDG seems to reside in a very low-density DM halo \citep[see also][]{kong22}, which might have implications for its stability, which we will investigate in this study. 

\citetalias{mancera24} discussed how the model of Case 2 seems to be a more realistic representation of the DM halo of $\rm{AGC 114905}$, given that a baryon fraction larger than the cosmological value, as in Case 1, is hard to justify physically (see their Section 5.2). We will come back to this point in Section~\ref{discussionUDGs}, but for the rest of the paper we will investigate equally both models, in order to see whether their behaviour is different, at least from a stability point of view.

\section{Methods}\label{methods}
The scope of this work is to develop numerical simulations (using the code AREPO, see below) of idealised galaxies whose properties resemble those of the observations outlined above, with the final goal of investigating whether the galaxy $\rm{AGC\ 114905}$ is stable against global gravitational instabilities. In this Section, we describe in detail how we build the initial configurations of our models and we outline the main features of our numerical experiments.

   \begin{figure}
   \includegraphics[clip, trim={0.5cm 1.5cm 1.5cm 3.8cm}, width=\linewidth]{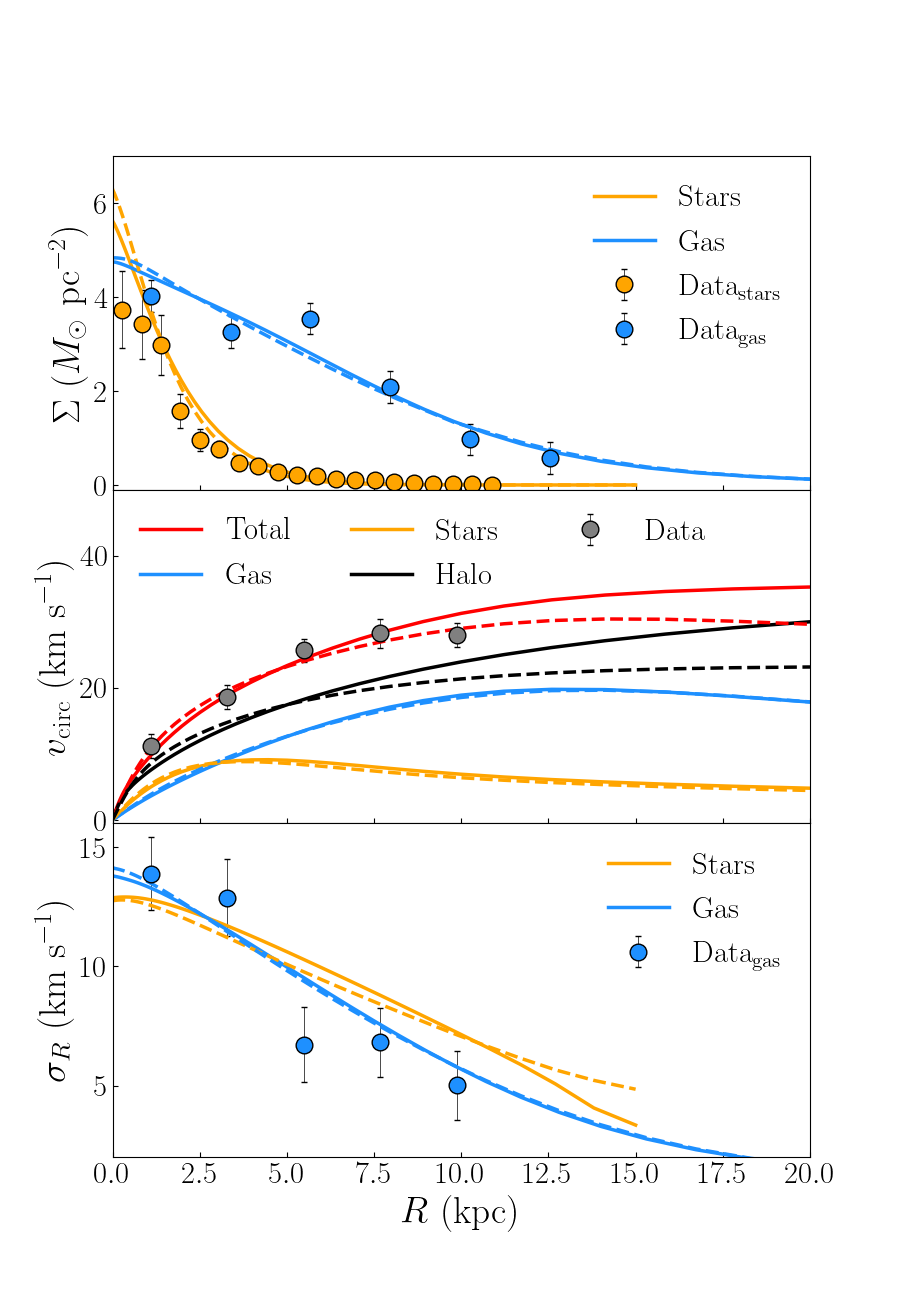}
   \caption{Comparison between the observational data points from \citetalias{mancera24} (see Section~\ref{data}) and the initial conditions of our two main models (Case 1 depicted by dashed curves and Case 2 by solid curves, respectively) produced with AGAMA (see Section~\ref{methods}). Properties of the stellar disc are shown in orange and of the gas disc in blue. Top panel, surface density profiles; Central panel, total circular velocities of the system (model in red and data in gray), with respective contributions from the baryonic and the DM (black curves) components; bottom panel, radial component of the velocity dispersion (the other two components have very similar values and are not shown for clarity). Models and data are (by construction) in good agreement with each other. Note that the gas surface density profile is slightly more extended than the rotation curve and the velocity dispersion, since these require a higher signal to noise ratio. The small amount of gas at $R>10$ kpc is not enough to constrain the HI kinematics.}
              \label{fig:ICagama}%
    \end{figure}

\subsection{Initial conditions}\label{section2}

We build the initial conditions of our idealised simulations through a distribution function (DF) based approach, using the Action-based GAlaxy Modelling Architecture library \citep[AGAMA,][]{vasiliev19}. While the following approach is strictly valid only for a collisionless system, we will extend it to a system that includes also a disc of gas, as we will describe in detail in Section~\ref{setup}.

A collisionless system with a sufficiently large number of particles can be fully described by a one-particle distribution function, which in a steady state depends only on the integrals of motion, in the approach of AGAMA, actions {\bf\emph{J}}. We first create a system of three components, two discs (one resembling the stellar and one the `gas' disc) and one spherical DM halo, each of them specified by its DF. AGAMA allows the user the create different types of DF with different methods, for both disky and spherical components. 

\subsubsection{Dark matter halo}\label{DMic}
We assume that the DM halo is described by a \textsc{core}NFW profile (see equation~\ref{eq:coreNFW}) and we fix its parameters to the values found by \citetalias{mancera24} for Case 1 and Case 2 (see Section~\ref{rotcurvedec}), investigating both cases separately. Once the DM density profile and its corresponding potential are defined, we use AGAMA to build an ergodic  distribution function with the Eddington inversion formula \citep{eddington16}. Given that in a spherical potential the energy is a function of the actions, this DF is action-based \citep[see e.g.][]{posti15} and is defined as `quasi-isotropic' in AGAMA. It can indeed be used also with a potential $\Phi(r)$ that is not necessarily the one giving rise to the original DM density profile and can also be not spherical.

\subsubsection{Disc components}
As for the disc components, we assume two quasi-isothermal DFs in AGAMA. In order to create such a DF, one needs to assume the values of the three input functions 
$\widetilde{\Sigma}(R)$,
$\widetilde{\sigma}_{R}(R)$ and $\widetilde{\sigma}_{z}(R)$. Note that these functions are defined for $z=0$ (midplane) and are only used to build the DF, from which the actual surface density and velocity dispersion profiles can later be generated, without being necessarily the same as the input ones \citep[see][their equation 20 and Section 4.3]{vasiliev19}.

For simplicity, we assume an input disc surface density given by an exponential form:
\begin{ceqn}
\begin{equation}\label{eq:discSurf}
\widetilde{\Sigma}_i(R) = \Sigma_{i,0} \exp(-R/R_{\rm{d},i})\ ,
\end{equation}
\end{ceqn}
where $\Sigma_{i,0}$ and $R_{\rm{d},i}$ are the central surface densities and scale radii of the $i$-th component (stars or gas). We also assume an exponential vertical profile, so that
\begin{ceqn}
\begin{equation}\label{eq:discdens}
\widetilde{\rho}_i(R,z) = \widetilde{\Sigma}_i(R) \times \frac{1}{2h} \exp(-|z|/h_i)\ ,
\end{equation}
\end{ceqn}
where $h_i$ is the scale height of the $i$-th component. We finally define the input vertical and radial velocity dispersions of both discs to be exponentially declining with radius:
\begin{ceqn}
\begin{equation}\label{eq:discdisp}
\widetilde{\sigma}_{X,i}(R) = \sigma_{X,i,0} \exp(-R/R_{\sigma,X,i})\ ,
\end{equation}
\end{ceqn}
with $X=R,z$. We tune the values of the free parameters in equations~\eqref{eq:discSurf} -- \eqref{eq:discdisp} in order to obtain final density ($\Sigma(R)$) and velocity dispersion (the three components of $\sigma(R,z)$) profiles for both stars and gas in agreement with our observational constraints (Section~\ref{data}), as we explain below. 

\subsubsection{Self consistent model}\label{SCM}

Given a number of components, AGAMA allows to build a self consistent, equilibrium model, by iteratively solving a system of three equations until convergence: i) the mapping of the actions {\bf\emph{J}}$(x,v)$, which depend on the potential $\Phi$ and is done through the St{\"a}ckel fudge method \citep{binney12}; ii) the density of each component, which is given by the integral of the respective DF; iii) the Poisson equation that relates the total density and potential. We apply this method on our three-component system, obtaining converged models after approximately 5 iterations. 
We use a trial and error approach in order to choose the best values of the parameters defining the distribution functions, to find a converged model that is the closest possible to the data, focusing in particular on the observed surface density profiles of stars and gas, on the circular velocity (extracted from the HI rotation curve) and on the HI velocity dispersion. The stellar velocity dispersion is unknown from the observations. Therefore, we choose a profile that would most closely resemble the gas velocity dispersion in order to have a comparable scale-height between gas and stars. The adopted value is also similar to what is found in nearby gas-rich dwarf galaxies \citep[e.g. Fig 2 in][and references therein]{mancera21}. We have tested that using a lower stellar velocity dispersion (and therefore a lower stellar scale-height) would not impact the results of this work (see also Section~\ref{comparison}).

We repeat the same procedure for both DM halos described in Section~\ref{DMic}. The final profiles of stars and gas for both models (dashed lines for Case 1 and solid lines for Case 2), as well as their comparison with the data, are shown in Figure~\ref{fig:ICagama}. By construction, both models lead to densities and kinematic properties that are consistent with the observations. For clarity, we report only one component of the gas and star velocity dispersion (specifically the radial component $\sigma_R$): the other two components have indeed very similar values, as we construct models with velocity dispersions close to isotropic (see Appendix~\ref{dispComponents} for more details).

Note that the profiles obtained with the iterative procedure explained above do not have an analytical form and do not exactly coincide with the surface density profiles used by \citetalias{mancera24} for the stellar and the gas disc (see Section~\ref{rotcurvedec}). Indeed, rather than reproducing exactly their results, we focus here on creating models that are self consistent and in an initial equilibrium state (which does not mean that the galaxy will be also \textit{stable} against perturbations), with the only requirement of being in agreement with the available observational data. The total masses of the three components (calculated with AGAMA by integrating the 3d density profiles) for the two different models are consistent with the estimates from \citetalias{mancera24} and are shown in Table~\ref{tab:inCond}\footnote{Table~\ref{tab:inCond} shows the DM mass within $2r_{\rm{vir}}$, while $\log{M_{200}/M_{\odot}}=9.59, 10.13$ for respectively Case 1 and Case 2. These values are slightly higher, but consistent within the uncertainties with the estimates of \citetalias{mancera24}. The differences are due to the creation of the self-consitent model with AGAMA, which tends to slightly modify the initial input DM density profile.}.

{ 
 \begin{table}
\begin{center}
\begin{tabular}{*{5}{c}}
(1)&(2)&(3)&(4)&(5)\\
\hline  
\hline
 Model & Component & $\log \frac{M_{\rm{tot}}}{M_{\odot}}$ & $\#_{\rm{part}}$ & $ m_{\rm{part}}/M_{\odot}$ \\
\hline 
\multirow{3}{4em}{Case 1} & Stars & $7.97$ & $10^5$ & $927$\\
 &Gas & $9.1$ & $1.3\times10^6$& $974$\\
 & Dark Matter & $9.9$ & $8\times10^6$& $993$\\
 \hline
\multirow{3}{4em}{Case 2} & Stars & $8.0$ & $10^5$& $1071$\\
 & Gas & $9.1$ & $1.3\times10^6$& $971$\\
 & Dark Matter & $10.57$ & $3.5\times10^7$& $1016$\\

\hline
\end{tabular}
\end{center}
\captionsetup{}
\caption[]{Properties of the initial N-body realizations of each component for Case 1 and Case 2, which are used to create the initial conditions of the simulations both with and without hydrodynamics. (1) model name; (2) type of component; (3) total mass of the respective component (note that the total DM mass is $M_{\rm{DM}}(<2r_{\rm{vir}})$, within which we extracted the particle realization); (4) number of particles used for each component; (5) particle mass.}\label{tab:inCond}
   \end{table}
   }

\subsection{Numerical simulations}\label{numsims}
We run our simulations using the moving-mesh, magnetohydrodynamical N-body code AREPO \citep{springel10,weinberger20}, a multiporpose code that has been used for a large variety of astrophysical problems, from cosmological simulations of galaxy formation and evolution \citep[e.g.][]{vogelsberger14} to accretion disks \citep[][]{fiacconi18}. The gravitational forces between the particles in the simulation domain are calculated using a TreePM algorithm \citep[][]{bagla02} and the time integration is performed through a second-order leapfrog scheme. The (magneto) hydrodynamics equations are instead solved by discretizing the fluid quantities on a moving, adaptive Voronoi mesh. The fluxes between the cells are calculated using an exact Riemann solver and the fluid quantities are evolved with a second-order finite-volume scheme \citep[][]{pakmor16}. For additional details on the implementation and the functionalities of AREPO, see \cite{weinberger20}.

In this work, we use AREPO to run idealised simulations of the UDG AGC 114905, starting from the initial conditions of the two models presented in Section~\ref{rotcurvedec} (Case 1 and Case 2). For both models we run two different simulations: we first assume that the models are entirely collisionless, so that also the gas disc is treated in the same way as the stars and the dark matter halo; we then include also the hydrodynamics, without including cooling, star formation and feedback (or any other type of baryon physics) as their implementation would be outside the scope of this work. While the second approach is more appropriate, we explore both cases in order to have a more direct comparison with the work of \citetalias{sellwood22}.

The first step necessary to run the simulations is to extract N-body realizations of the three components that constitute our self-consistent galaxy model. This can be done in AGAMA, by sampling the DFs with $N$ particles using an adaptive multidimensional rejection sampling algorithm. The number of particles used for each component and the corresponding particle masses (approximately $10^3\ M_{\odot}$) are reported in Table~\ref{tab:inCond}. Given that the total mass of a halo described by equation~\eqref{eq:coreNFW} would be infinite, we define a truncation radius $r_t = 2r_{\rm{vir}}$, where $r_{\rm{vir}}$ is the virial radius, which sets the initial boundaries of the dark matter halo, over which we extract the distribution of DM particles. Note that this approach is different from the one used by \citetalias{sellwood22}, who assumed for the DM halo a static contribution to the gravitational potential, while we utilize a live halo that will evolve with our simulations (see also Appendix~\ref{darkmatter}). Earlier work has shown this might favour the formation of a central bar, making the galaxy less stable than in the static case \citep[e.g.][]{athanassoula02,sellwood16}.

The combination of the N-body distributions of the three components corresponds to the initial snapshot that we can directly use to start the collisionless simulations in AREPO. In all cases (also for the simulations including the hydrodynamics, see below) we adopt a 3-dimensional box with a size of $400$ kpc and a gravitational softening length of $8$ pc, following the procedure of \cite{dehnen11}. In particular the softening length is set to be equal to the mean value of the stellar and gas softening, both obtained by dividing the respective system's size (which we take to be equal to the disc scale length\footnote{The final density profiles shown in Figure~\ref{fig:ICagama} do not have an analytical form, therefore we fit directly the observed surface density profiles of stars and gas with an exponential profile and we use the best-fit disc scale length to estimate an appropriate softening length. We have anyway tested that varying this value within a factor of two does not impact the evolution of our simulated galaxies.}) by the square root of the respective number of particles.

   \begin{figure*}
   \includegraphics[clip, trim={2.5cm 1cm 2.5cm 1.5cm}, width=\linewidth]{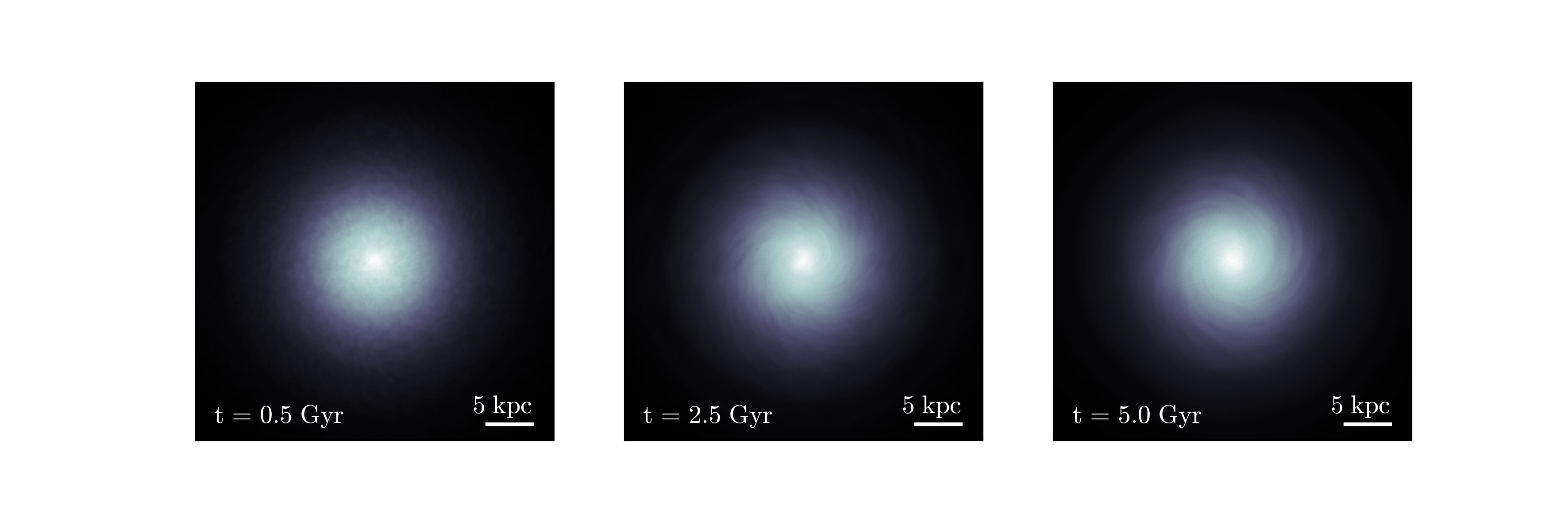}
   \includegraphics[clip, trim={2.5cm 1.5cm 2.5cm 1cm}, width=\linewidth]{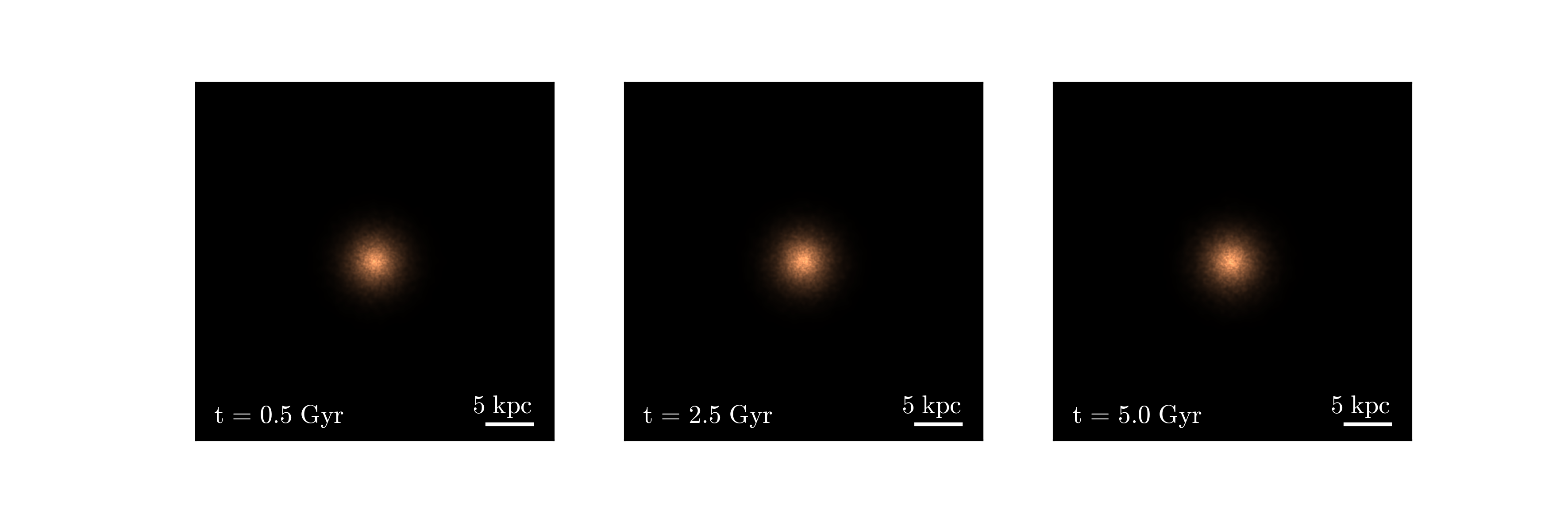}
   \caption{Zoom-in on the face-on density distributions of the gas (top panels) and the stellar (bottom panels) discs of the Case 2 model with hydrodynamics included, at three different times in the simulation evolution ($t=0.5,2.5,5$ Gyr). Both discs do not seem to develop global gravitational instabilities and their structure remains essentially unaltered with time.}
              \label{fig:snapshots}%
    \end{figure*}

\subsubsection{IC for the hydrodynamical simulations}\label{setup}
In order to include the gas treatment, we need to convert our N-body realizations of the gas particles obtained with AGAMA into appropriate initial conditions \citep[see also][for other examples]{deg19,teppergarcia24}. To this purpose, we first assume that the gas obeys the ideal gas law:
\begin{ceqn}
\begin{equation}\label{eq:eos}
P_{\rm{gas}} = (\gamma - 1) \rho_{\rm{gas}} U_{\rm{int}} ,
\end{equation}
\end{ceqn}
where $\gamma=5/3$ is the politropic index for an adiabatic gas and $U_{\rm{int}}$ is the gas specific internal energy. In our approach, we define this quantity as:
\begin{ceqn}
\begin{equation}\label{eq:uint}
U_{\rm{int}}(R,z) = \frac{\sigma^2_z(R,z)}{(\gamma -1)} (1 - \beta^2)\ ,
\end{equation}
\end{ceqn}
where $\sigma_z(R,z)$ is the gas vertical velocity dispersion obtained with AGAMA and $\beta$ is an arbitrary factor that defines how much of the vertical support against the gravitational field is given by internal energy (the rest will be due to the actual particle velocity dispersion). We also modified the initial velocities of the gas particles in order to take into account the additional support along the radial direction given by the pressure gradient, according to the equation \citep[see][]{springel05}:
\begin{ceqn}
\begin{equation}\label{eq:vphi}
v^2_{\phi, \rm{gas}} = R \left( \frac{\partial \Phi}{\partial R} + \frac{1}{\rho_{\rm{gas}}} \frac{\partial P_{\rm{gas}}}{\partial R}\right) .
\end{equation}
\end{ceqn}

The final step is to make our initial conditions suitable for the grid-based calculations that AREPO uses for the hydrodynamical quantities. To this purpose, AREPO allows to generate a grid with properties that follow as closely as possible the initial gas particle distribution, extended with an additional coarse background grid that fills the entire computational domain \citep[see][for more details]{weinberger20}. We apply this procedure to the $N_{\rm{gas}}$ particles describing the gas disc. We note that creating the grid tends to lower the original velocity dispersion of the particles created with AGAMA, while leaving the average rotational and radial velocities unchanged. We checked that the ratio between the velocity dispersion after and before introducing the grid is equal to about 0.27, without any dependence with the radius $R$. Hence, we assume that the factor $\beta$ in equation~\eqref{eq:uint} is equal to 0.27, so that the `remaining' dispersion increases the internal energy $U_{\rm{int}}$. This procedure ensures that the total velocity dispersion, given by the sum in quadrature of the internal and the particle dispersions (which effectively represent respectively the gas thermal and turbulent dispersion), is equal to the original dispersion of the particles in the N-body distribution created with AGAMA and in turn to the observed velocity dispersion of the gas. The above procedure defines the initial conditions for our hydrodynamical simulations. We additionally assume an adaptive softening length for the gas particle with a minimum value of $4$ pc (half of the value of the gravitational softening length of collisionless particles). Finally, we point out that we use a refinement scheme in AREPO that keeps the mass of the gas cells approximately constant during the evolution of the simulation (within a factor 2 from the particle masses specified in Table~\ref{tab:inCond}).

   \begin{figure*}
   \includegraphics[clip, trim={0cm 0cm 0cm 0cm}, width=\linewidth]{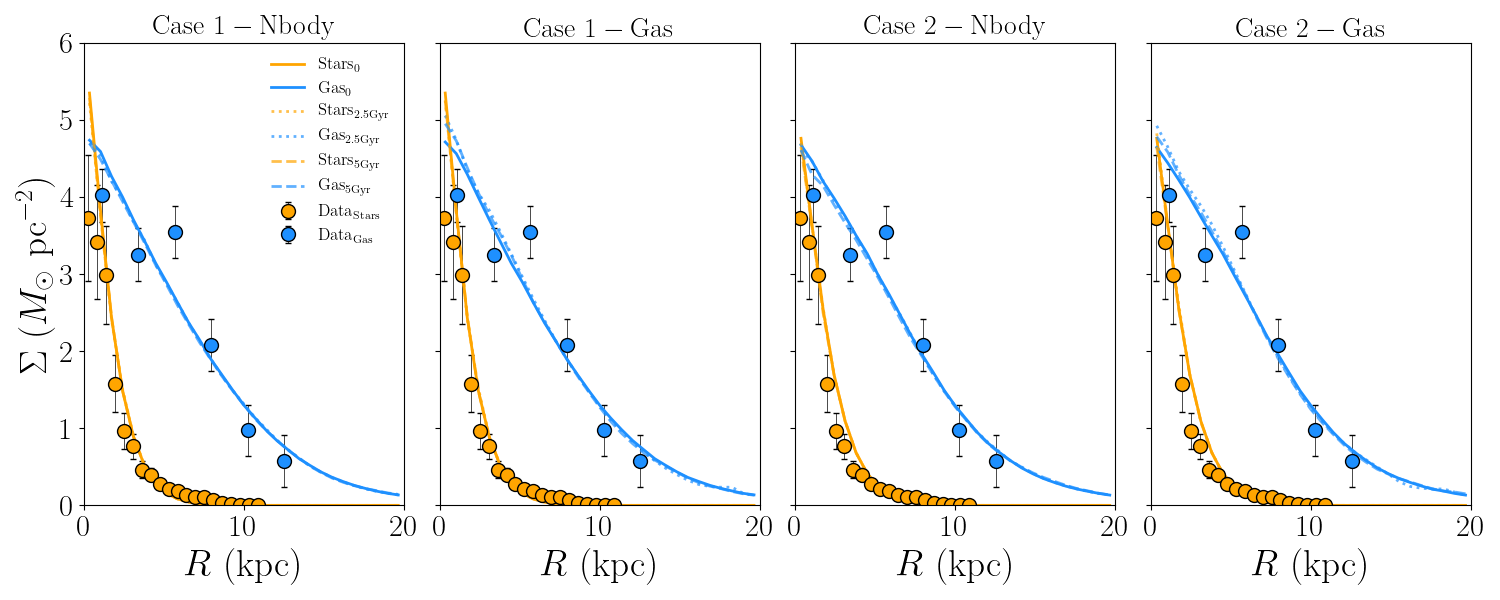}
   \caption{1-dimensional surface density profiles of the stellar (orange) and gas (blue) discs in the four main simulations analyzed in this work (`Nbody' columns are for the collisionless simulations, while `Gas' columns are for the simulations with hydrodynamics), at $t=0,2.5,5$ Gyr. The profiles are calculated by dividing the disc into concentric rings wit $\Delta R \simeq 0.7$ kpc. The points show the corresponding observational data from \citetalias{mancera24}. All the simulated profiles are consistent with the data and do not exhibit any significant evolution with time.}
              \label{fig:profiles}%
    \end{figure*}
    
\section{Results}\label{results}
We evolve the idealised galaxy models described above (see Sections~\ref{section2} and~\ref{numsims}) for 5 Gyr, in order to investigate whether they would develop or not global gravitational instabilities at any point during the course of the simulation. We will explore in this Section the results of our 4 main simulations, i.e. the models resulting from the choice of Case 1 and Case 2 for the DM halo (see Sections \ref{rotcurvedec} and~\ref{DMic}), both with and without the inclusion of the hydrodynamics.

\subsection{Density distribution}
\subsubsection{Face-on view}
In order to first qualitatively assess the stability of our galaxy model, we show in Figure~\ref{fig:snapshots} three snapshots at three different times ($t=0.5, 2.5, 5$ Gyr) of the 2D density distributions of the gas (top) and stellar (bottom) discs, seen face-on, for the simulation of Case 2 with hydrodynamics included. The snapshots for the other three cases do not show any significant differences and therefore we do not include them here, but we encourage the reader to visit \href{https://doi.org/10.5281/zenodo.14852154}{this link}, where full videos of the gas and stellar discs of all the simulations run for this study can be found.

What is immediately apparent by looking at the snapshots is that both discs do not develop any instabilities and maintain their initial shape up until the end of the simulation. There is a slight hint of spiral arm features appearing in the gas disc towards the end of the simulation (such features are not present in the collisionless case, but the difference is minimal), but the overall distribution remains unchanged and the discs appear overall stable. This result can be seen also more quantitatively in the four panels of Figure~\ref{fig:profiles}, where we plot the surface density of the two discs (gas in blue and stars in orange), as a function of the cylindrical radius $R$, at the beginning, at $2.5$ Gyr and at the end of the four simulations. For each snapshot, the profiles are calculated by computing the surface density over concentric rings with $\Delta R \simeq 0.7$ kpc. We can see how in all cases both profiles evolve very slightly (if at all) up until the end of the simulation. In the cases with hydrodynamics (second and fourth panel) the gas surface density slightly increases in the central regions, but this change does not seem significant and the evolution is entirely consistent with the collisionless cases and, most importantly, with the observational data. The above results therefore point towards the main conclusion of this work: the UDG AGC 114905 is stable and does not develop global gravitational instabilities for several Gyr, contrary to what previously found in \citetalias{sellwood22}.

\subsubsection{Edge-on view}
\begin{figure*}
   \includegraphics[clip, trim={2.5cm 1cm 2.5cm 1.5cm}, width=\linewidth]{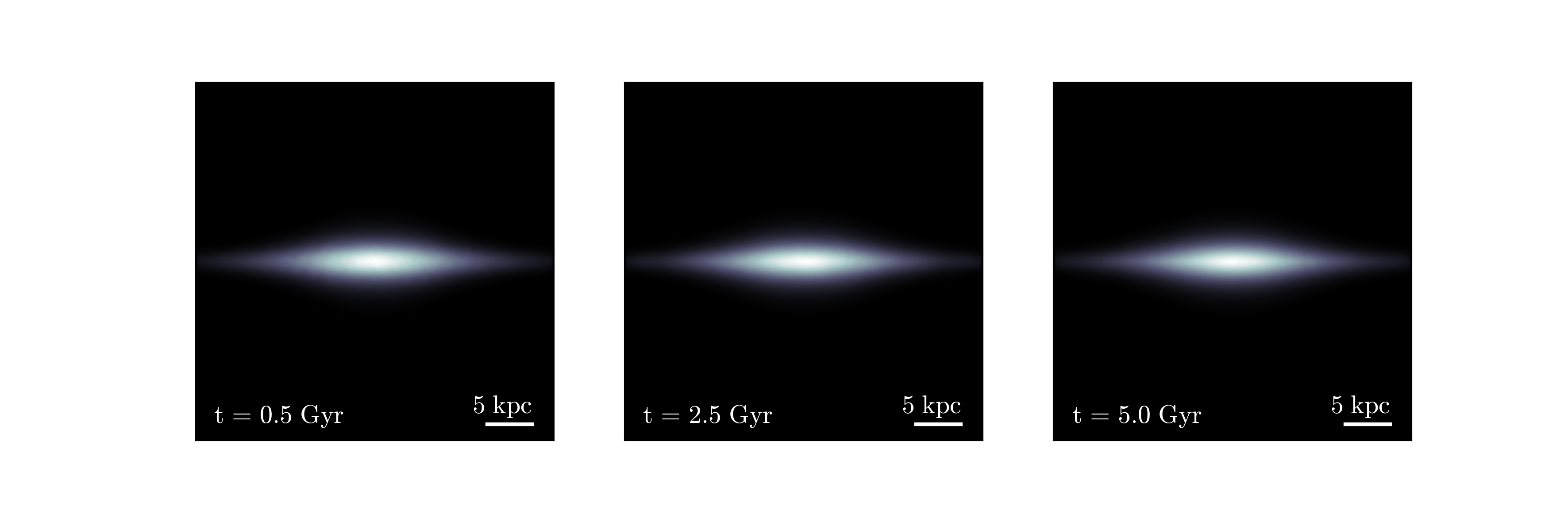}
   \includegraphics[clip, trim={2.5cm 1.5cm 2.5cm 1cm}, width=\linewidth]{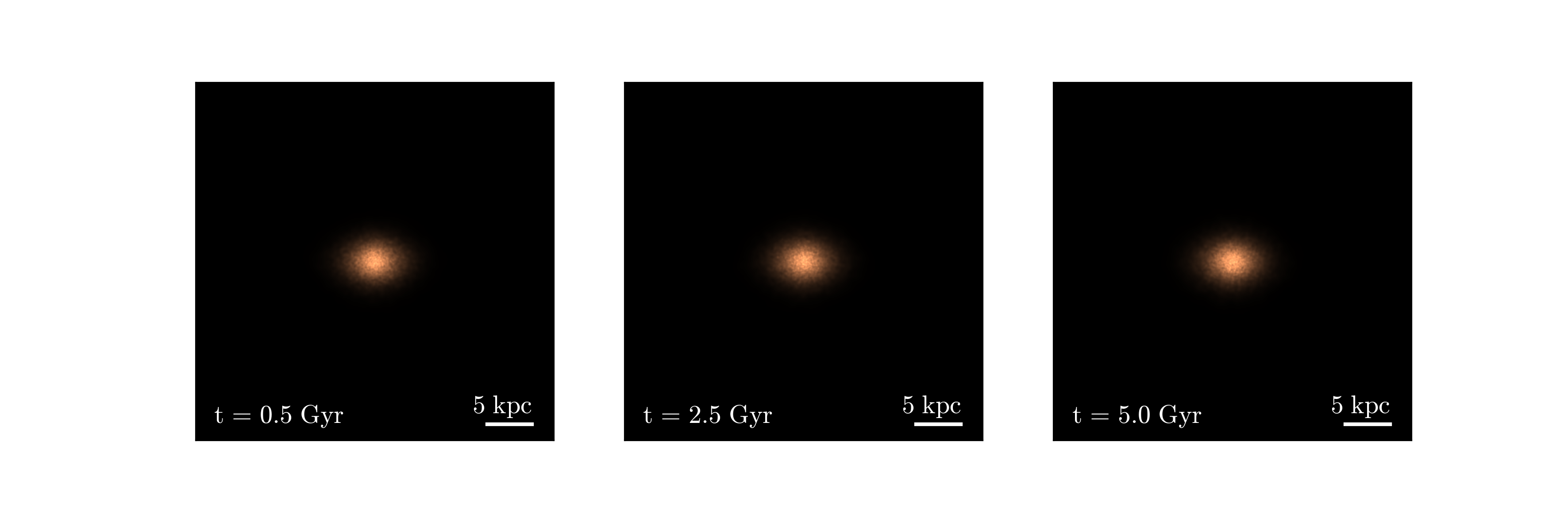}
   \caption{Same as Figure~\ref{fig:snapshots}, but for the gas (top) and stellar (bottom) discs seen edge-on.} \label{fig:snapshots_edgeon}
    \end{figure*}

\begin{figure*}
   \includegraphics[clip, trim={0cm 0cm 0cm 0cm}, width=\linewidth]{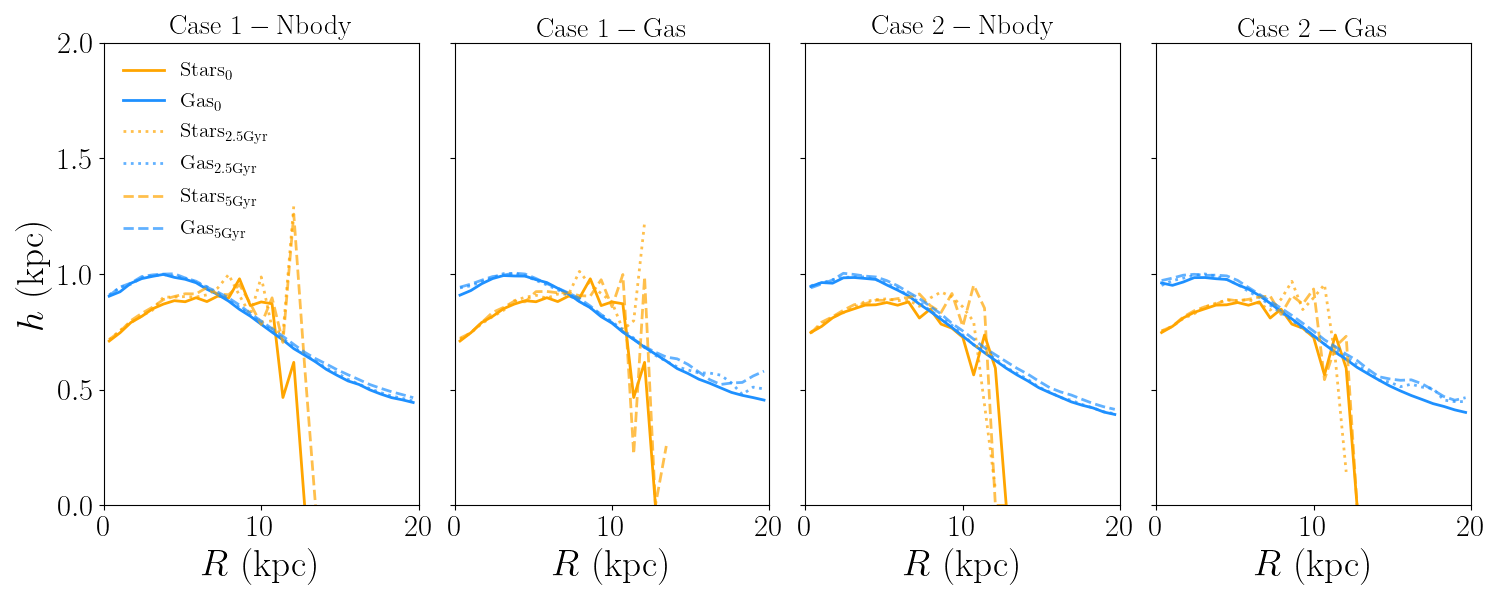}
   \caption{Similar to Figure~\ref{fig:profiles}, but for the stellar (orange) and gas (blue) scale-height profiles obtained at $t=0,2.5,5$ Gyr from the 4 main simulations analyzed in this study.}
              \label{fig:scaleHeights}%
   \end{figure*}

In Figure~\ref{fig:snapshots_edgeon} we show the same snapshots of Figure~\ref{fig:snapshots}, seen from an edge-on view. Also from this view, it is immediately clear how the discs do not evolve significantly with time, implying that the galaxy is not affected by global instabilities and it maintains its initial gas and stellar distributions. More quantitatively, the vertical shape of the two discs is visible in Figure~\ref{fig:scaleHeights}. Here, we plot the scale height of the discs in the four simulations, by calculating, for each ring over which we computed the surface density shown in Figure~\ref{fig:profiles}, the standard deviation of the $z$ distribution of the gas and star particles\footnote{In the case of the gas disc in the simulations with hydrodynamics, this quantity (and all the other quantities involving averages) is weighted by the mass of the gas cell. Indeed, while in the initial N-body distribution all the particles have the same mass, generating the grid for the gas initial conditions (see Section~\ref{setup}) slightly modifies the mass distribution. We anyway exclude cells with masses significantly lower than (less than 1$\%$) the values reported in Table~\ref{tab:inCond}. Most of these excluded cells belong to the background grid and are located far from the center of the computational box (where the galaxy is).}. Looking at the gas profiles (blue curves), we can see how the disc reaches a maximum of about $1$ kpc before declining at $R\gtrsim5$ kpc, reaching scale heights of about $0.5$ kpc in its external regions. By construction, the stellar disc also exhibits a similar trend in its scale height (orange profiles), given that we built models in which stars and gas have a similar velocity dispersion profile. Note that the stellar thickness quickly drops to zero at $R\approx12$ kpc, which represents the outer edge of the stellar disc: no star particles are present in our simulations beyond this radius. In all simulations, both stars and gas show little to no evolution in their scale height profiles between the start and the end of the simulation. The only noticeable differences are present at $R\gtrsim 7$ kpc for the stellar discs and at $R\gtrsim 15$ kpc for the gas disc, where the densities are very low and very few particles determine the scale heights of the two discs. 

Note that the vertical disc structure has a peculiar shape: while discs are expected to be flared \citep[e.g.][]{romeo92,yim14,marasco17,bacchini19,mancera22flaring} as an effect of vertical hydrostatic equilibrium, with the scale height increasing going to larger $R$, the scale height of this simulated gas disc remains approximately constant up to a few kpc and then declines outwards. Even though this behaviour is not common, it is consistent with previous estimates of the thickness of the HI disc of AGC 114905 based on hydrostatic equilibrium (\citealt{bacchini24}, \citetalias{mancera24}). The reason is likely due to the drastic drop of the velocity dispersion in the outer parts, leading to a pressure not strong enough to support the flaring.

      \begin{figure*}
   \includegraphics[clip, trim={0cm 0cm 0cm 0cm}, width=\linewidth]{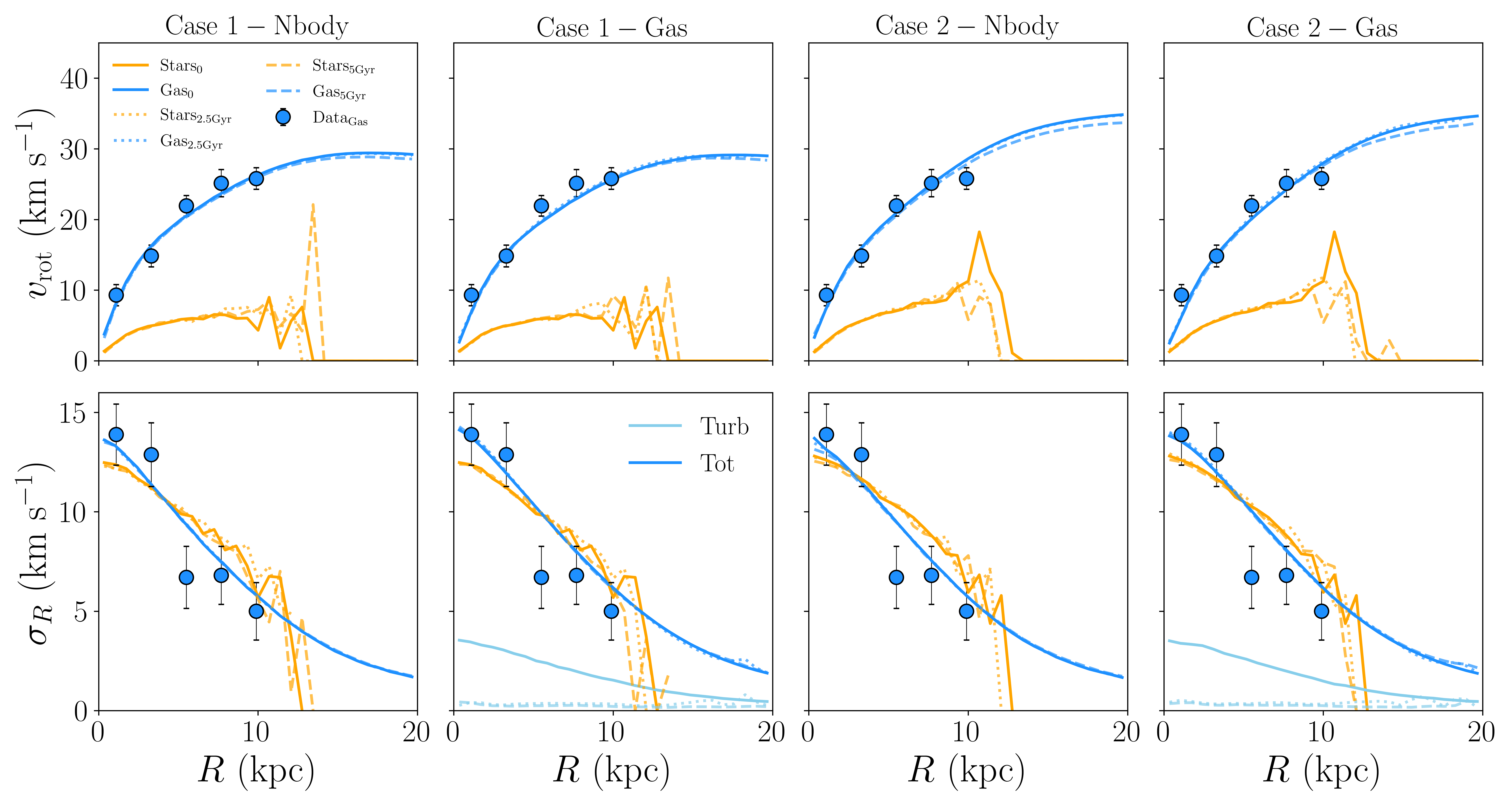}
   \caption{Kinematic properties of the stellar (orange) and the gas (blue) discs of the 4 main simulations analyzed in this study, with the corresponding data-points from \citetalias{mancera24}. The top panels show the rotational velocities and the bottom panels the radial component of the velocity dispersion. For the two cases with hydrodynamics (second and fourth column), we show both the turbulent (light blue) and the total (darker blue, thermal plus turbulent) velocity dispersion. Note that the most external rings where the stellar kinematics is calculated, which exhibit a very noisy behaviour, contain a negligible amount of particles.}
              \label{fig:kinematics}%
   \end{figure*}

\subsection{Disc kinematics}
We have seen above that both the vertical and radial density distributions do not change during the course of the simulation. In this Section we focus instead on the kinematics of both the gas and the stellar disc, in order to further confirm whether the galaxy is stable or not. Figure~\ref{fig:kinematics} shows the rotational velocities (top panels) and the velocity dispersions (bottom panels) of gas (blue curves) and stars (orange), for the same simulations and at the same times analyzed above. All the profiles are extracted by taking the average quantities of the particles belonging to the same rings utilized for the plots in Figures~\ref{fig:profiles} and~\ref{fig:scaleHeights}. 

From the top panels of Figure~\ref{fig:kinematics}, we can see how the gas rotates initially (by construction) at the same speed of the observations and how this rotational motion does not change as the simulation evolves, again indicating stability (note that in the simulations with hydrodynamics in the second and fourth columns, the gas disc has a slightly lower rotational velocity compared to the collisionless cases, due to the additional correction of the radial pressure gradient, see equation~\ref{eq:vphi}). The only sign of evolution can be seen at $R\gtrsim10$ kpc, where the gas rotational velocity is $1$ or $2$ km s$^{-1}$ lower by the end of the simulation. This is due to the evolution of the DM halo, as we discuss more in detail in Appendix~\ref{darkmatter}. This effect is however negligible and does not affect at all the region of the galaxy where observational constraints are available.
As for the stellar disc, there is to date no observational evidence of its kinematics in AGC 114905, but we can see how in all cases the simulated stellar discs also rotate with the same velocity during the course of the simulation (except for the very external rings that contain a negligible amount of particles) and therefore appear stable.

The top panels of Figure~\ref{fig:kinematics} also show how the rotation of the stellar disc is more strongly affected by the asymmetric drift (see equation~\ref{eq:adrift}), given that the slope of the stellar surface density profile is much steeper with respect to that of the gas disc. The stellar component tends therefore to rotate much slower than the gas: this prediction of our model, if true, implies that i) the stellar kinematics of these galaxies will be very challenging to measure, given the very low rotational velocities, in current and future observational surveys, and ii) that even if a low rotational velocity is observed in the stellar kinematics of a dwarf galaxy, the gas disc could still rotate at a significantly higher speed.

The bottom panels of Figure~\ref{fig:kinematics} show instead the gas and stellar velocity dispersion at the beginning and at the end of the simulation. In the collisionless cases, we can see how the velocity dispersion (for clarity we show again here the radial component $\sigma_{\rm{R}}$ only, but the other two components are very similar, as shown in Appendix~\ref{dispComponents}, and therefore any of these components can be used to compare with the observations) exhibits almost no evolution during the 5 Gyr of the simulation. For the simulations with hydrodynamics (second and fourth columns) we plot separately the total velocity dispersion (blue), given by the sum in quadrature of thermal and turbulent motions, and the particle (or turbulent) velocity dispersion (light-blue) of the gas disc\footnote{As for the collisionless cases, we show here only the radial component of the velocity dispersion of the cells, but the other components have the same trend (and the same is valid for the total dispersion, given that the internal dispersion is isotropic).}. We can see how the total velocity dispersion is constant with time and is always consistent with the observational constraints, while the turbulent dispersion, that since the start of the simulation is not dominant, decreases with time getting to values very close to zero by the end of the simulation. This is because, being the gas a collisional component (and since we are in the absence of radiative cooling), the kinetic energy due to turbulence is quickly dissipated into internal energy. We have checked that, using different values of $\beta$ (see equation~\ref{eq:uint}), the kinetic, turbulent energy of the gas particles is almost completely dissipated into internal, thermal energy within 1 Gyr, with the simulations showing very similar results. Finally, the orange profiles show how also the stars maintain the same velocity dispersion throughout the simulation.

To conclude, the (absence of) evolution of both the rotational velocity and the velocity dispersion confirms how AGC 114905 does not develop global gravitational instabilities throughout its evolution, assuming any of the two DM halos (Case 1 and Case 2) explored in this work.

\section{Discussion}\label{discussion}
\subsection{Comparison with previous work}\label{comparison}
The main point of comparison for the present study is the recent work of \citetalias{sellwood22}. As already discussed in Section~\ref{intro}, using similar simulations to those presented here, they found that the gas and stellar discs of AGC 114905 are not globally stable, contrary to our findings. Below we discuss the main differences and similarities between our works, in addition to those already addressed in the previous sections.

   \begin{figure}
   \includegraphics[clip, trim={1cm 1cm 1cm 1cm}, width=0.49\linewidth]{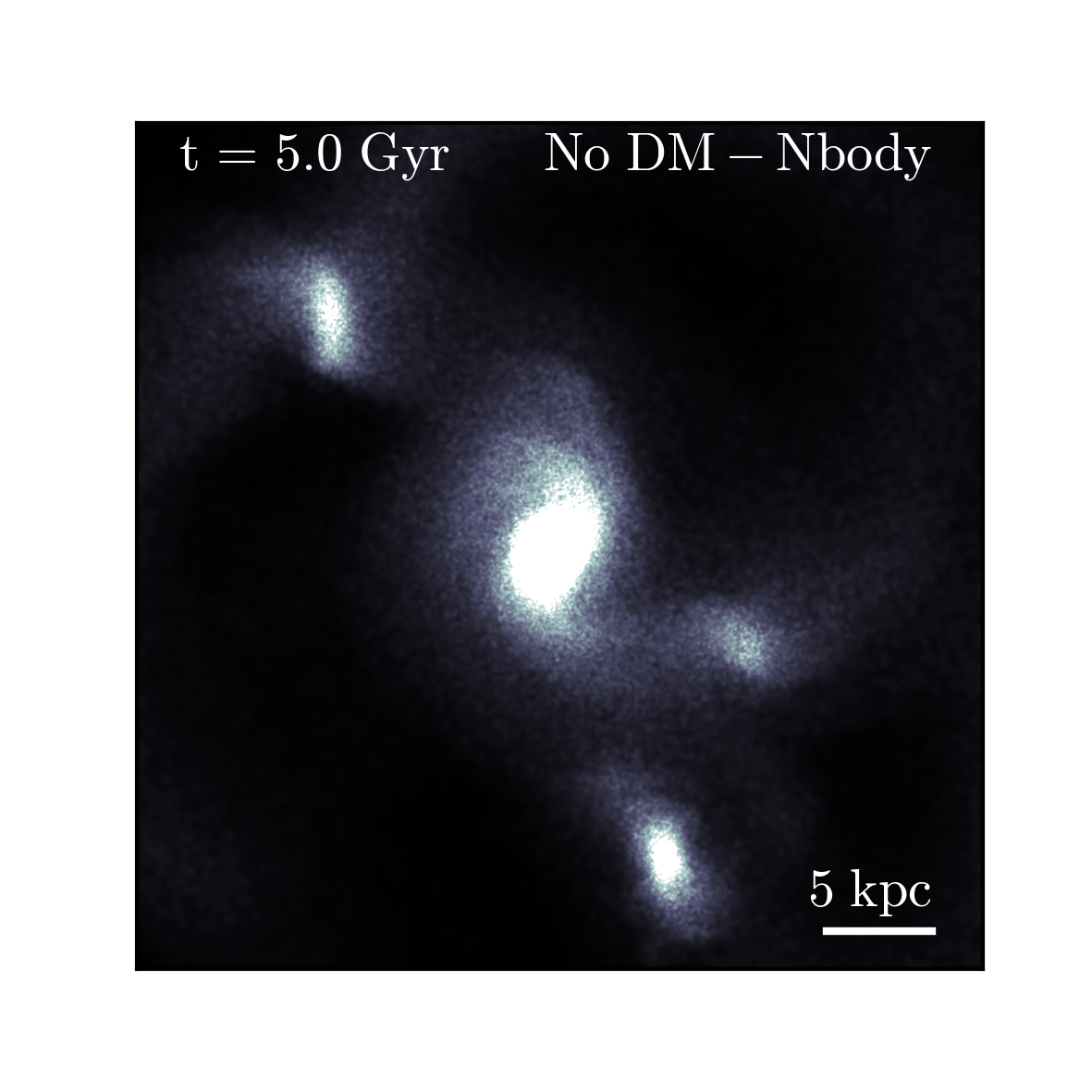}
   \includegraphics[clip, trim={1cm 1cm 1cm 1cm}, width=0.49\linewidth]{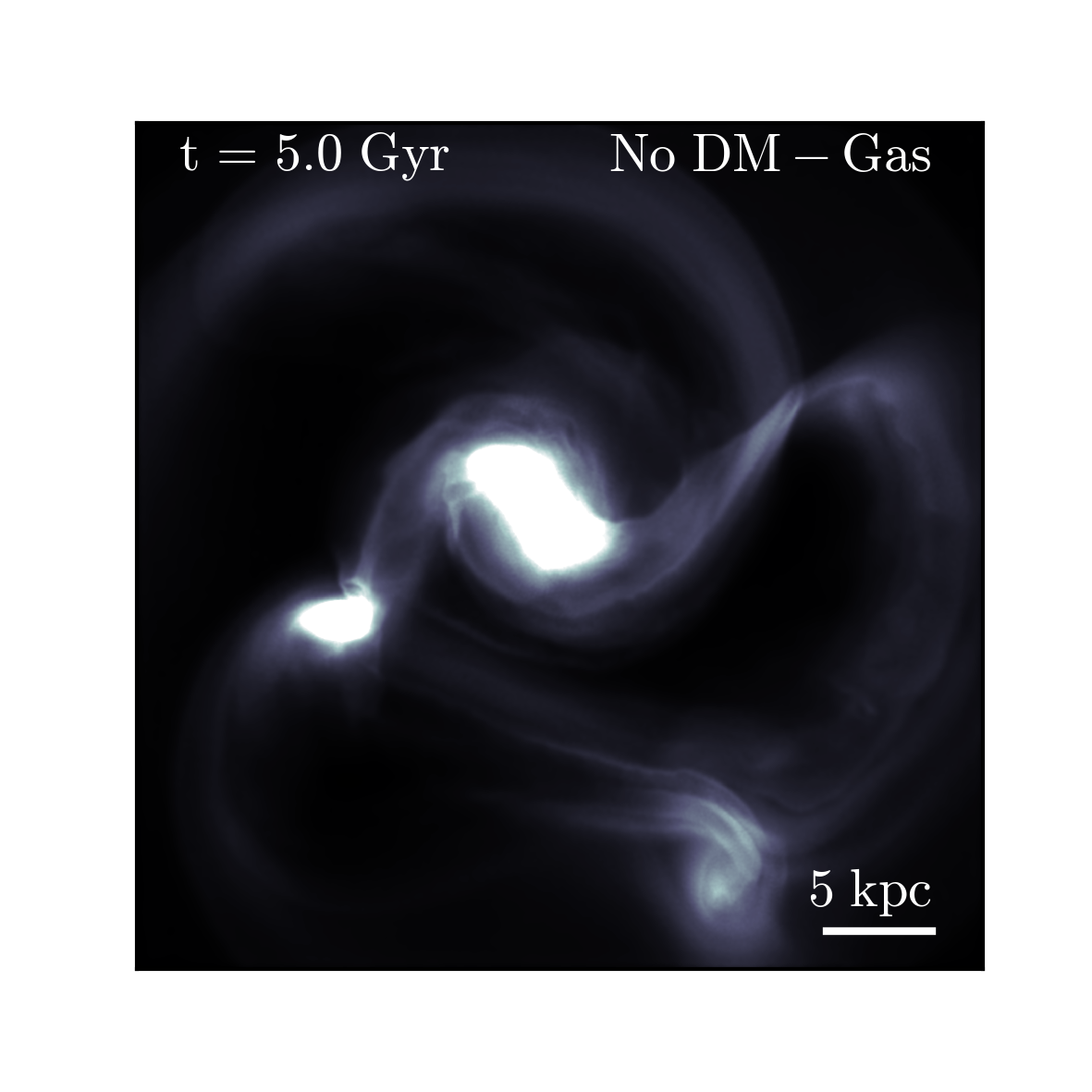}      \includegraphics[clip, trim={1cm 1cm 1cm 1cm}, width=0.49\linewidth]{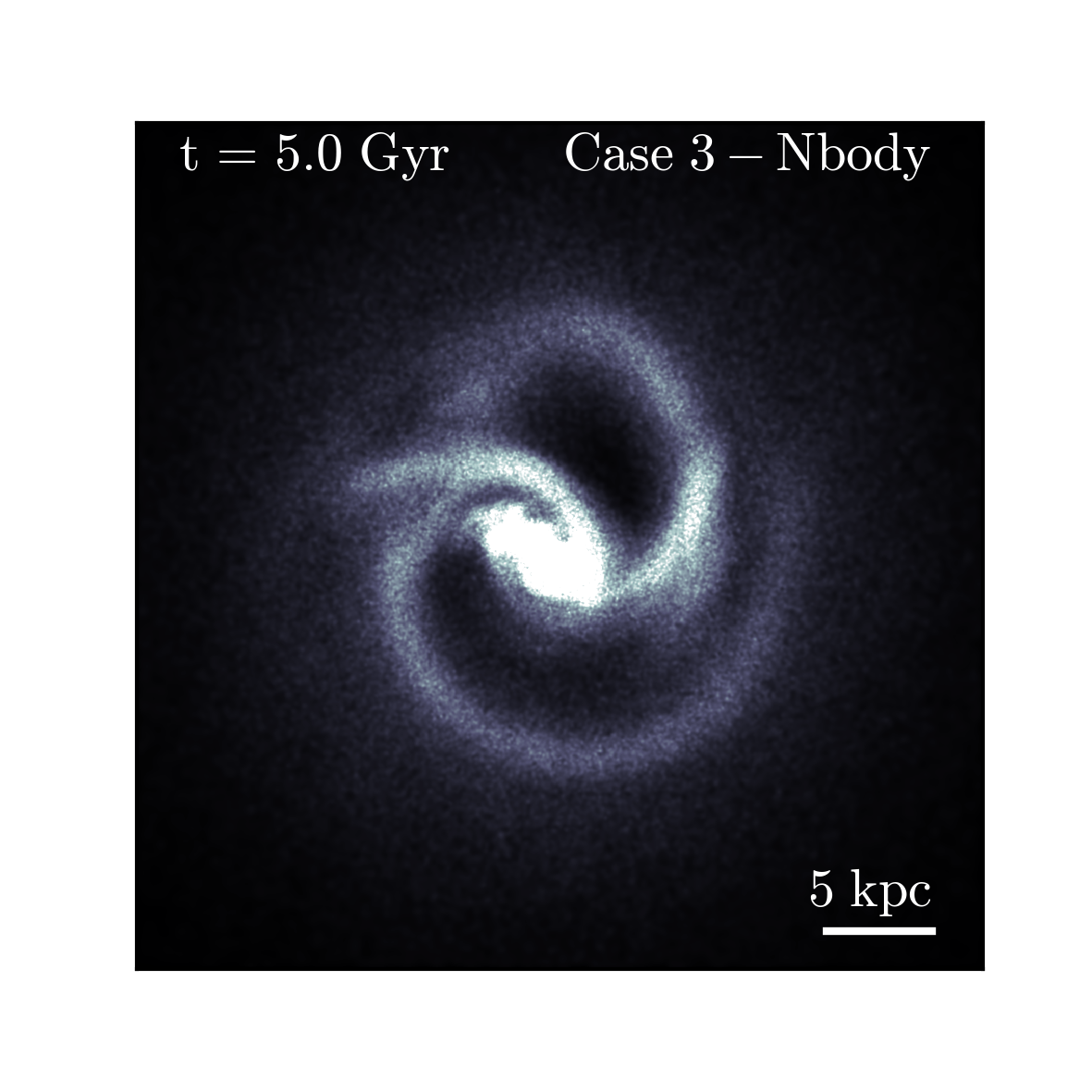}
    \hspace{0.005cm}
   \includegraphics[clip, trim={1cm 1cm 1cm 1cm}, width=0.49\linewidth]{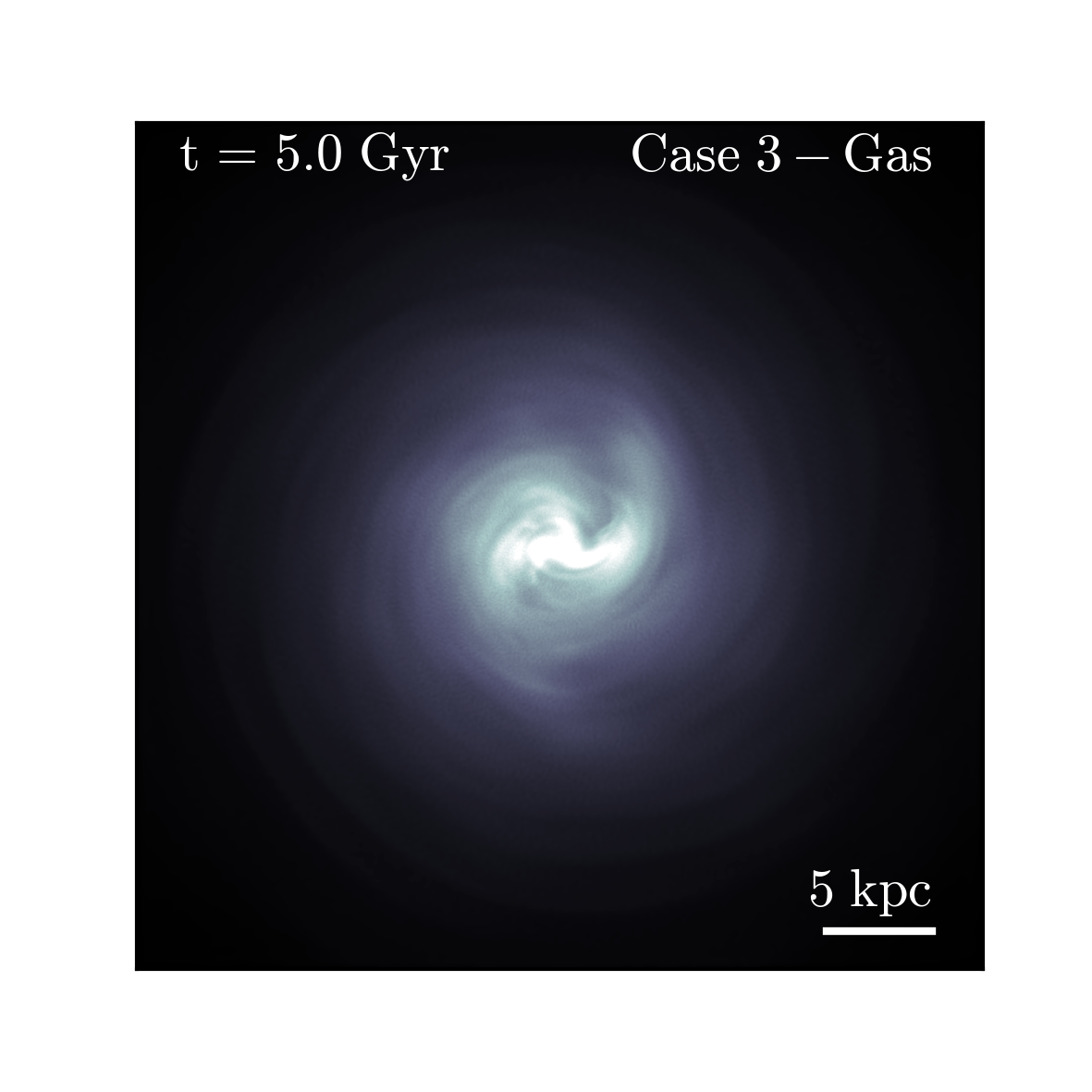}
   
   \caption{Face-on views of the gas disc at the end of the simulation for the 4 models discussed in Section~\ref{comparison} (see the main text for more details). Top panels: `No DM' models, where the adopted DM halo has a significantly lower mass with respect to our fiducial cases and the velocity dispersion is approximately $5$ km s$^{-1}$. Bottom panels: `Case 3' models, where the DM halo is the same as in `Case 2' and the velocity dispersion is again $5$ km s$^{-1}$. Three out of the four simulations develop global gravitational instabilities that dramatically deform the initial disc structure. This is in agreement with the results of \citetalias{sellwood22}.} \label{fig:snapshots_discussion}
    \end{figure}

Differently from this study, \citetalias{sellwood22} based their simulations on the data of \cite{mancera22}, the only available at the time. This results in slight discrepancies between our surface density profiles of gas and stars (which should however not dramatically change the evolution and the stability of the UDG) and most importantly in different circular velocities and velocity dispersions. First, the lower (about 15-20\% with respect to the values of \citetalias{mancera24}) circular velocities found by \cite{mancera22}, led \citetalias{sellwood22} to simulate a system composed by only two discs and no spherical DM halo, or with a DM halo with a lower mass compared to what we utilize here: within $10$ kpc, the halo of \citetalias{sellwood22} contains about $6.6\times10^8\ M_{\odot}$, while we use $M_{\rm{DM}}(<10\ {\rm kpc})\approx9.2\times10^8\ M_{\odot}$ and $M_{\rm{DM}}(<10\ {\rm kpc})\approx1.1\times10^9\ M_{\odot}$ for respectively our Case 1 and Case 2 (see also Table~\ref{tab:resultsInst}). Additionally, they then employed a velocity dispersion of about $5$ km s$^{-1}$ in both discs, consistently with the findings of \cite{mancera22}, while we use a velocity dispersion that goes from about $14$ km s$^{-1}$ in the center down to about $5$ km s$^{-1}$ at $R=10$ kpc, following the updated observational constraints of \citetalias{mancera24} (see Figure~\ref{fig:ICagama}). We attribute the difference in our findings mainly to the different dark matter halo and velocity dispersion employed in our model.

\begin{figure}
   \includegraphics[clip, trim={0.5cm 1.5cm 1.6cm 3.5cm}, width=\linewidth]{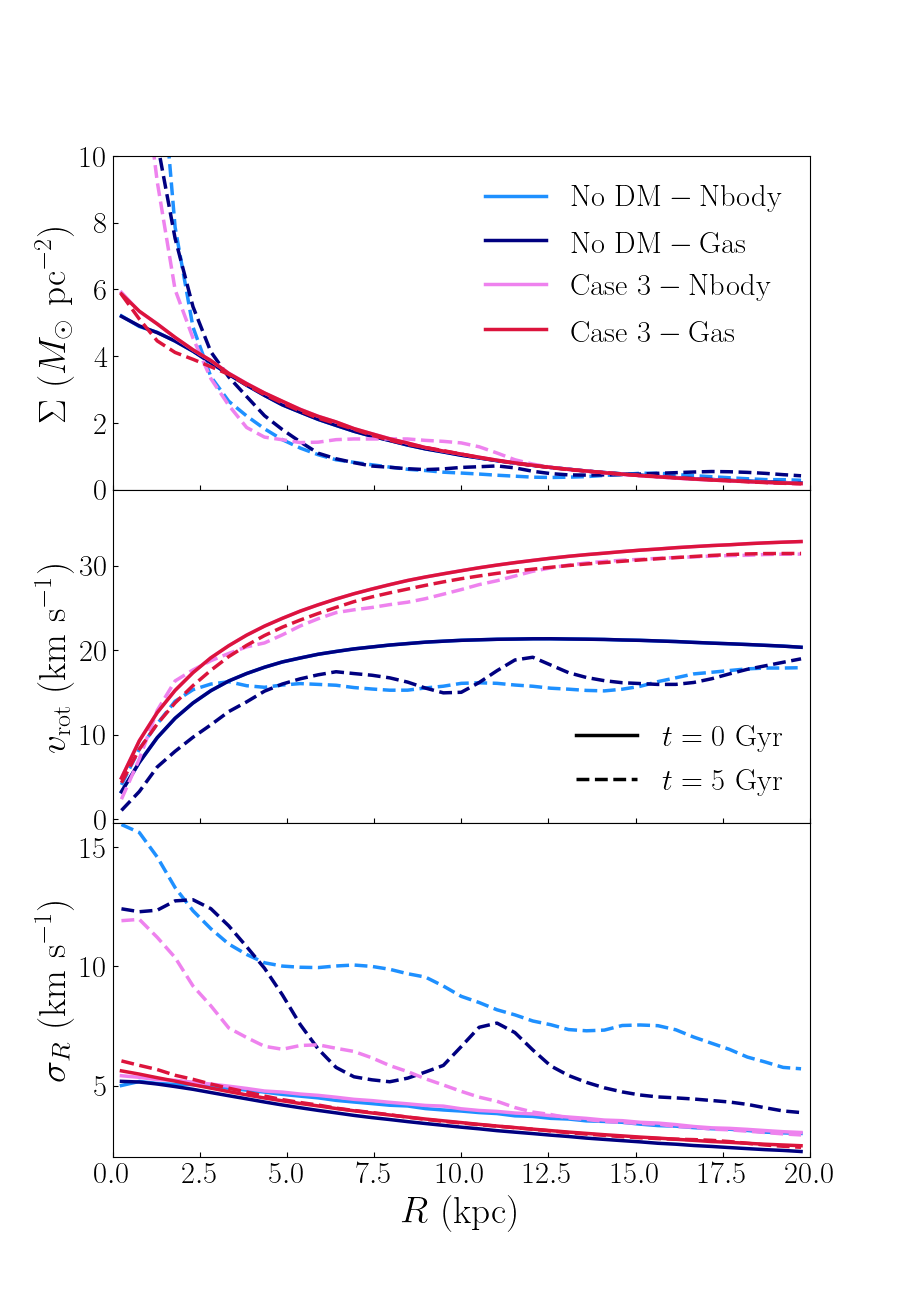} 
   \caption{Properties of the gas disc in the four models presented in Section~\ref{comparison} at the beginning (solid curves) and at the end (dashed curves) of the simulations. Surface densities on top, rotational velocities in the central panel and velocity dispersions at the bottom. Only the `Case 3' simulation with hydrodynamics remains stable for 5 Gyr, while in the other three cases all three properties drastically change due to global instabilities, in agreement with the results of \citetalias{sellwood22}. Note that the rotational velocities of the `No DM' models (as well as the velocity dispersions of all the models presented in this figure) are too low to be consistent with the data reported in \citetalias{mancera24}, but would be in agreement within the uncertainties with the previous estimates from \protect\cite{mancera22}.} \label{fig:profiles_discussion}
    \end{figure}

To demonstrate the above statement, we run an additional set of simulations, modifying the velocity dispersion and the dark matter halo in order to be more comparable with those used by \citetalias{sellwood22}. To compare with their first simulation, where they investigated the stability of AGC 114905 in the absence of a DM halo, we design an initial equilibrium system with AGAMA assuming a DM halo with a very low mass ($M_{\rm{DM}}(<2r_{\rm{vir}})\approx 2\times10^9\ M_{\odot}$ and $M_{\rm{DM}}(<10 {\rm kpc})\approx 2.3\times10^8\ M_{\odot}$), so that the baryonic component largely dominates the gravitational potential everywhere. We additionally lower the gas velocity dispersion (and also the stellar one accordingly) to $5$ km s$^{-1}$, roughly constant with radius. We simulate such system (which we call `No DM') twice, both with and without the inclusion of the hydrodynamical effects for the gas disc and we let the simulations run for $5$ Gyr, as in our fiducial cases. The results of such experiments are shown in Figures~\ref{fig:snapshots_discussion} (top panels) and \ref{fig:profiles_discussion} (cyan and blue curves). Both from the density snapshots and from the radial profiles of surface density, rotational velocity and velocity dispersion we can see that, after $5$ Gyr, the system has developed strong gravitational instabilities that completely destroyed its initial shape. These results are in excellent agreement with those of \citetalias{sellwood22}. The evolution of the gas discs is also similar (see the full videos at \href{https://doi.org/10.5281/zenodo.14852154}{this link}) to what previously found, with the disc rapidly developing a central depression and then being fragmented in multiple structures. The inclusion of the hydrodynamics seems to affect the evolution of the discs, but the two simulations produce qualitatively similar results. We finally estimate the time when the instability seems roughly to first take place, which we arbitrarily define as the time $t_{\rm{inst}}$ when the average relative difference between the initial and the current radial profile of at least one of the properties of the gas disc ($\Sigma$, $v_{\rm{rot}}$ and $\sigma_R$) is larger than 20$\%$ (this time generally also agrees well with the time when major deviations from the initial gas distribution appear in the 2-d videos of the simulations). We find that in both cases $t_{\rm{inst}}\approx0.7$ Gyr (see also Table~\ref{tab:resultsInst}), meaning that the instabilities develop very quickly and the galaxy is largely unstable. As already discussed by \citetalias{sellwood22}, we conclude that AGC 114905 needs a significant contribution from the DM halo (with a dark matter mass within 10 kpc of at least $\approx10^9\ M_{\odot}$) in order to be stable. Note that we do not show the observational data points in Figure~\ref{fig:profiles_discussion}, as the rotational velocities and especially the velocity dispersions of this model would be too low and not consistent with the data from \citetalias{mancera24}.

{ 
 \begin{table}
\begin{center}
\begin{tabular}{*{6}{c}}
(1)&(2)&(3)&(4)&(5)&(6)\\
\hline  
\hline
 \vspace{0.1cm}
 Sim. Id &
 $M_{\rm{DM}}(< 10 {\rm{kpc}})$
 & $<\sigma>$ & $t_{\rm{inst}}$ & $<Q>$ &  $\mathcal{E}$\\
 &$M_{\odot}$& km s$^{-1}$& Gyr&&\\
\hline 
Case 1 - Nbody &
$9.2\times10^8$& 9.5 & - & 1.95 & 1.12\\
Case 1 - Gas &
$9.2\times10^8$& 9.5 & - &  2.08 & 1.12\\
Case 2 - Nbody &
$1.1\times10^9$& 9.5 & -  & 1.97 & 1.31\\
Case 2 - Gas &
$1.1\times10^9$& 9.5 & -  & 2.10 & 1.31\\
No DM - Nbody &
$2.3\times10^8$& 4.5 & $0.7$  & 0.85 & 0.77\\
No DM - Gas &
$2.3\times10^8 $& 4.5 & $0.7$ & 0.91 & 0.77\\
Case 3 - Nbody &
$1.1\times10^9$& 4.6 & $3.2$  & 1.11 & 1.28\\
Case 3 - Gas &
$1.1\times10^9$& 4.6 & - &  1.19 & 1.28\\

\hline
\end{tabular}
\end{center}
\captionsetup{}
\caption[]{Summary of the main properties related to the stability of $\rm{AGC\ 114905}$, for all the simulations performed in this study. (1) Simulation Id; (2) dark matter halo mass within 10 kpc (if the DM halo is not massive enough the disc tends to be unstable); (3) average (radial) velocity dispersion for $1<R<10$ kpc in the gas disc (if the disc is not sufficiently dinamically hot it tends to be unstable); (4) time in which the global instability starts to develop (hot discs in equilibrium with a sufficiently massive DM halo do not develop global instabilities); (5) average Toomre parameter $Q$ for $1<R<10$ kpc within the gas disc (models with $Q>1$ are expected to be stable against local instabilities, see \citealt{safronov60,toomre64}); (6) $\mathcal{E}$ parameter (collisionless and collisional models with respectively $\mathcal{E}<1.1$ and $\mathcal{E}<0.9$ are expected to develop a global bar instability, see \citealt{efstathiou82, Christodoulou95}).}\label{tab:resultsInst}
   \end{table}
   }

As an additional test, we create a system where we lower the velocity dispersion of gas and stars to $5$ km s$^{-1}$, using the dark matter halo of our Case 2 model, in order to investigate how much the velocity dispersion alone is responsible for the different outcomes between our simulations and those from \citetalias{sellwood22}. Such a model (which we once again simulated for $5$ Gyr, with and without hydrodynamical effects) is directly comparable with the `Case 3' model explored in \citetalias{sellwood22}, where they employed a more massive DM halo, more similar to the one adopted here. The collisionless case (see bottom left panel of Figure~\ref{fig:snapshots_discussion} and violet profiles in Figure~\ref{fig:profiles_discussion}) confirms the results of \citetalias{sellwood22}, as we find that the gas disc (and similarly the stellar disc) develops global instabilities (with $t_{\rm{isnt}}\approx3.2$ Gyr) that leave the galaxy strongly perturbed by the end of the simulation, although in a less dramatic fashion with respect to the case with a very small amount of dark matter. This result proves that, as expected (see Section~\ref{discussionInstability}), if the stellar and the gas discs were dynamically too cold ($\sigma\lesssim 5$ km s$^{-1}$) they would tend to develop global instabilities within a few Gyr. Such a low velocity dispersion seems however to be incompatible with the observations of \citetalias{mancera24}\footnote{We checked that models with intermediate velocity dispersions ($\sigma\sim 8$ km s$^{-1}$), potentially in agreement within the uncertainties with the observational data, are instead stable, as previously found also by \citetalias{sellwood22}.}. 

Interestingly, the `Case 3' simulation with hydrodynamics seems instead to not show strong deviations from its initial configuration (see bottom right panel of Figure~\ref{fig:snapshots_discussion} and red curves in Figure~\ref{fig:profiles_discussion}): as for our fiducial cases shown in Section~\ref{results}, none of the gas properties differ, at any time in the simulation, by more than $5\%$ from their initial values. We therefore mark this model as stable. While this may sound surprising, one possible explanation of this result is that the gas collisional effects have a stabilizing influence on this particular galaxy model. For example, using a linear analysis, \cite{rafikov01} found that a system composed by two gas collisional discs has a larger stable region compared to a system with a stellar (collisionless) and a gas disc, implying that collisional effects might have an effect in damping local instabilities. Analytical works on global instabilities (\citealt{efstathiou82,Christodoulou95}, see also discussion below) have also shown that stellar discs seem to be more susceptible to the development of bar instability with respect to gas discs. We stress that here we performed adiabatic simulations and the inclusion of baryon physics (using for example a model like SMUGGLE, \citealt{marinacci19}) would most likely affect the overall stability of a model so close to being unstable. The last result must therefore be taken with caution and more simulations will be needed to possibly confirm it. However, this shows how the various galaxy components must be properly modeled in order to accurately study the system's stability.

Overall we conclude that, if the DM halo is sufficiently massive and the gas and stellar discs sufficiently hot, the UDG AGC 114905 is stable against global instabilities and can survive unperturbed for several Gyrs, even for halos with extremely low concentrations. Models with colder discs and/or a significantly lower-mass DM halo could possibly lead the galaxy to be unstable (as in \citetalias{sellwood22}), but are excluded given that they have properties not in agreement with the latest observational data.

\subsubsection{Local and global instability}\label{discussionInstability}
As discussed by \citetalias{sellwood22}, global stability is likely related to the onset of local instabilities \citep[e.g.][]{lin64}: a disc that is locally unstable across the majority of its extension, would also be globally unstable. The most common parameter that has been used to investigate the local stability of a disc is the Toomre parameter \citep{toomre64}:
\begin{ceqn}
\begin{equation}\label{eq:Qpar}
Q = \frac{\kappa \sigma}{3.36 G\Sigma}\ ,
\end{equation}
\end{ceqn}
where $\kappa$ is the epicycle frequency, $G$ is the gravitational constant, $\sigma$ is the velocity dispersion along the radial direction and $\Sigma$ is the gas or stellar surface density. Note that equation~\eqref{eq:Qpar} is strictly valid for a stellar, collisionless disc, while a collisional fluid would require the factor $\pi$ instead of $3.36$ and the sound speed $c_{\rm{s}}$ instead of the velocity dispersion $\sigma$  \citep[see][]{safronov60,rafikov01,binney08}, but the differences are minimal. Regions of the disc with $Q<1$ are expected to grow local instabilities. We calculated this parameter for the gas disc (the stellar discs have always $Q>1$ due to the low stellar densities) in all the simulations performed in this work (using the factor $3.36$ for the simulations without hydrodynamics and $\pi$ for those with the inclusion of the hydrodynamical effects), in the region of the disc that is covered by observational data ($1<R<10$ kpc). The average value of $Q$ across this region is reported in Table~\ref{tab:resultsInst}. We find that all our fiducial simulations (Section~\ref{results}) have $Q$ that is largely higher than 1 everywhere in the disc, consistently with their stable evolution.
The simulations with a lower velocity dispersions have instead values much closer to 1, with the `No DM' cases (which have lower mass DM halos, implying lower values of $\kappa$) having $Q<1$, in agreement with our findings of instability shown above (Figures~\ref{fig:snapshots_discussion} and~\ref{fig:profiles_discussion}).

From the first definition more than 6 decades ago, more sophisticated versions of the $Q$ parameter have been developed, taking into account effects like the disc thickness \citep[e.g.][]{romeo92,romeo94}, which makes the disc more stable (as also found for $\rm{AGC\ 114905}$, see \citetalias{mancera24}), or the presence of multi-component systems \citep[e.g.][]{jog84,rafikov01,romeo13}. The most recent and accurate criterion is given by a 3-dimensional $Q$ parameter, $Q_{3D}$, that self-consistently takes into account the vertical structure of a disc in hydrostatic equilibrium with a certain gravitational potential \citep{nipoti23}, also derived for a system composed of two responsive thick disc components \citep{nipoti24}. As mentioned in Section~\ref{intro}, the single-component $Q_{3D}$ has in particular been calculated for the case of $\rm{AGC\ 114905}$, based on the observations of \citetalias{mancera24}, by \cite{bacchini24}. These authors found that $Q_{3D}>1$ everywhere in the disc, indicating stability, in excellent agreement with the findings of the present study.

The global instability of a disc has also been investigated for decades, both with numerical simulations~\citep[e.g.][]{hohl71,ostriker73} and linear perturbation analyses. \cite{toomre81} introduced a criterion for the onset of a global instability called swing amplification, a process that tends to increasingly amplify the disc's spiral patterns. He first introduced the parameter $X$, defined as:
\begin{ceqn}
\begin{equation}\label{eq:Xpar}
X = \frac{\kappa^2 R}{2\pi G\Sigma m}\ ,
\end{equation}
\end{ceqn}
where $m=2$. A disc is usually expected to be subjected to strong swing amplification factors if $X$ is not significantly larger than 1 and if $Q$ is close to unity, so that the disc is relatively stable, but responds efficiently to gravitational perturbations. We calculate the value of $X$ for the four models analyzed in this study (Case 1, Case 2, No DM and Case 3) and we find that in all our setups $1\lesssim X \lesssim 3$ almost everywhere for $1<R<10$\ kpc, without any significant differences between the various models. By combining these values with the $Q$ values reported in Table~\ref{tab:resultsInst}, our results seem to be in agreement with the expectations, with the models closer to local instability ($Q\sim1$) being also more influenced by the swing amplification instability (see the simulation snapshots in Figures~\ref{fig:snapshots} and \ref{fig:snapshots_discussion}).

Another type of global instability is the bar instability, which has been classically studied through the parameter $\mathcal{E}$, defined by \cite{efstathiou82} as:
\begin{ceqn}
\begin{equation}\label{eq:eln}
\mathcal{E} = \frac{v_{\rm{max}}}{(G M_{\rm{disc}}/R_{\rm{disc}})^{1/2}}\ ,
\end{equation}
\end{ceqn}
where $v_{\rm{max}}$ is the maximum rotational velocity, $M_{\rm{disc}}$ is the disc mass and $R_{\rm{disc}}$ is the disc scale length. We report the derived values of $\mathcal{E}$ for all our simulations in Table~\ref{tab:resultsInst}. \cite{efstathiou82} found that a bar instability will develop in stellar discs with $\mathcal{E}\lesssim 1.1$, while this criterion becomes $\mathcal{E}\lesssim 0.9$ for a gas disc \citep{Christodoulou95}. These predictions seem to work well for the majority of our simulations, given that the `Case 1' and `Case 2' models have $\mathcal{E}> 1.1$ and the `No DM' model, the only one strongly unstable, has $\mathcal{E}< 0.9$. However, while the slightly different criteria between a gas and a collisionless disc could explain the different outcomes of the two `Case 3' simulations with and without hydrodynamics, looking at the values reported in Table~\ref{tab:resultsInst} we can see that in principle we would expect both simulations to be stable. We conclude that, for galaxies that are potentially close to instability, such a criterion is likely too simplistic to determine whether or not a global instability (leading to the formation of a bar) would take place, as previously found by \cite{romeo23}. In these cases, numerical simulations, like those performed in this work, remain necessary in order to study disc stability.

\subsection{Implications for UDG formation}\label{discussionUDGs}
   \begin{figure}
   \includegraphics[clip, trim={0.5cm 1.5cm 1.6cm 3.cm}, width=\linewidth]{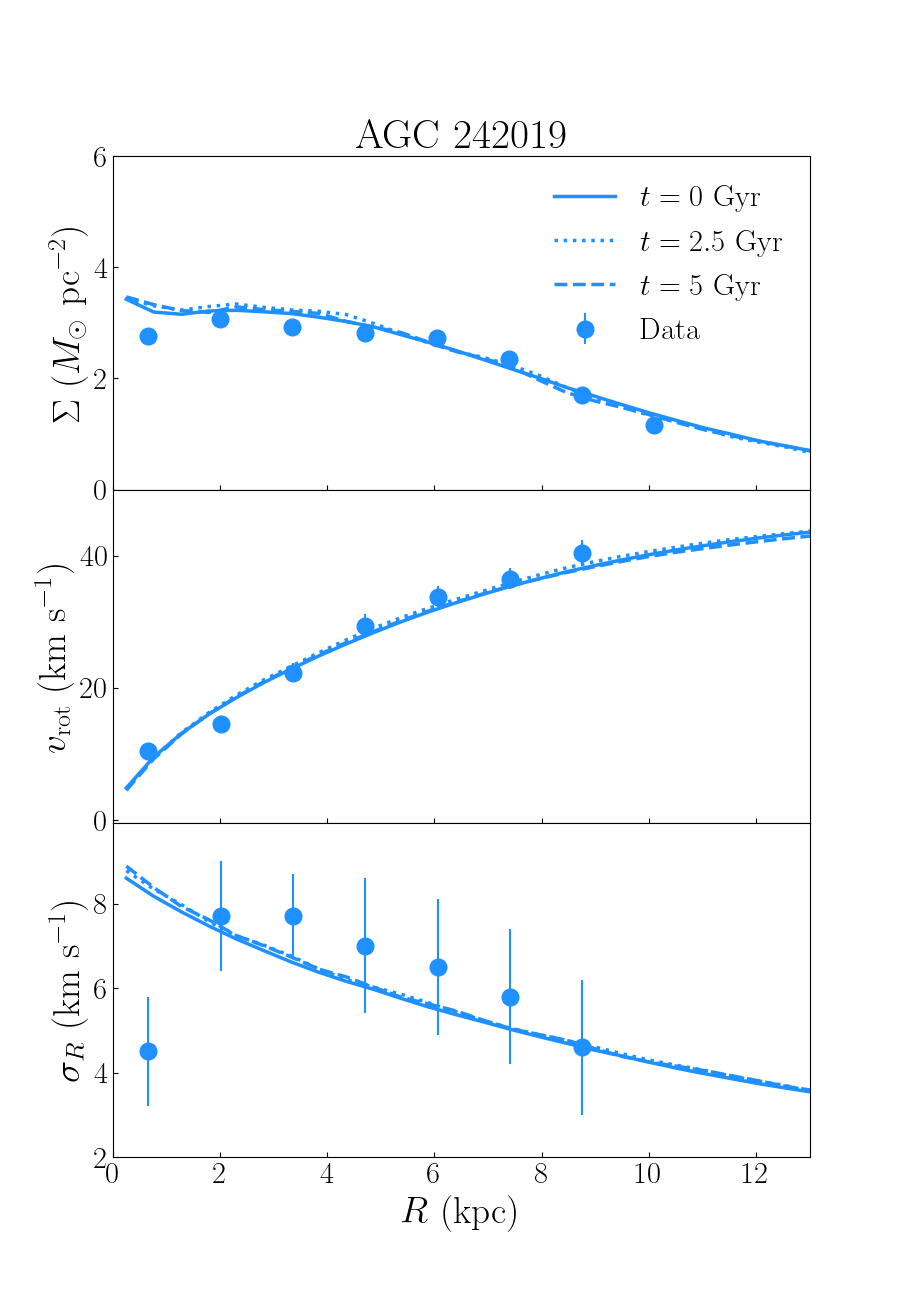} 
   \caption{Properties of the gas disc in the model presented in Section~\ref{discussionUDGs} for the UDG $\rm{AGC\ 242019}$, at the beginning (solid curves), at $2.5$ Gyr (dotted curves) and at the end (dashed curves) of the simulation. Surface densities on top, rotational velocities in the central panel and velocity dispersions at the bottom. The data points are taken from \protect\cite{shi21}. Except for the innermost (uncertain) estimate of the velocity dispersion in the bottom panel, the initial simulated profiles are (by construction) consistent with the observations. As for $\rm{AGC\ 114905}$, the simulated gas disc does not show any sign of instability (the same is true for the stellar disc) and does not significantly evolve with time. This result confirms that the findings of this paper can likely be applied to a broader class of gas-rich UDGs living in low-density DM halos.} \label{fig:shi}
    \end{figure}
The results of this paper are focused on a single galaxy, $\rm{AGC\ 114905}$. However, recently various HI-rich UDGs have been observed with DM halo properties similar to the galaxy analyzed here (see Section~\ref{intro}). \cite{kong22} found that the halos of seven UDGs (including $\rm{AGC\ 114905}$, being the most extreme of their sample) have lower concentrations with respect to what expected from CDM models. Of these galaxies, the two with the highest HI-data resolution are $\rm{AGC\ 114905}$ and $\rm{AGC\ 242019}$, first analyzed in \cite{shi21}. To follow up our findings, we additionally run one simulation whose initial conditions are calibrated on $\rm{AGC\ 242019}$. In particular, we create a 3-component system (stars, gas and DM), following the same procedure explained in Section~\ref{methods}\footnote{For this galaxy, to create the disky distribution functions we use as $\widetilde{\Sigma}(R)$ an exponential disc with a depression in the inner regions (see for example \citealt{oosterloo07}, \citetalias{mancera24}), rather than equation~\eqref{eq:discSurf}, given that the former produces a significantly better fit to the data.} to reproduce the observations of \cite{shi21} for the stellar and gas surface density, the HI rotational velocity and velocity dispersion (shown also as data-points in Figure~\ref{fig:shi}). For the DM halo, described again by equation~\eqref{eq:coreNFW}, we use the best-fit parameters found in \cite{kong22}. We also consider hydrodynamical effects in the same way explained in Section~\ref{setup}. In Figure~\ref{fig:shi}, we show the properties of the gas disc during the course of this simulation: as for $\rm{AGC\ 114905}$, we find that this galaxy (the stellar disc exhibits a similar behavior) does not evolve significantly and therefore is stable against global instabilities. This test, together with the fact that $\rm{AGC\ 114905}$ is the most extreme of the currently known HI-rich UDGs, points towards a generalization of the conclusion that gas-rich UDGs are generally stable.
 
Having confirmed that these UDGs can survive in atypical DM halos with very low concentrations, the main question remains how can these galaxies (and halos) form. \cite{kong22} found that halos with concentrations as low as those derived for their galaxy sample are present in the TNG50-1-Dark, dark-matter-only simulations from the IllustrisTNG project \citep{nelson19}, as a class of halos that formed with a high spin and at a late stage ($z\sim0.5$) of the Universe. Despite this similarity, \cite{kong22} also found that the derived central DM densities of both $\rm{AGC\ 114905}$ and $\rm{AGC\ 242019}$ are significantly lower than the corresponding simulated halos, highlighting the difficulty of producing such peculiar halos in a CDM Universe.

Accounting for baryon physics seems to be able to produce cored DM halos through stellar feedback \citep[e.g.][]{sales22}, but it is not clear whether such effects would be substantial enough in galaxies with very low star formation densities \citep[see e.g.][]{leisman17} and currently hosting a large amount of gas, with baryon fractions that are at least comparable to the cosmological average. While some hydrodynamical simulations, like NIHAO \citep{dicintio17} and FIRE \citep{chan18}, are able to reproduce gas-rich UDGs, the simulated halos in which these galaxies reside exhibit much larger densities with respect to the observations \citep[see][]{kong22}.

Given the struggles of CDM in reproducing these observations (even when considering feedback processes), alternative dark matter theories have been proposed to explain the formation of these halos (see \citetalias{mancera24}, for a more exhaustive discussion). Here we focus briefly on self-interacting dark matter \citep[SIDM, see][for a review]{tulin18}, which seems to be one of the most promising candidates. In SIDM, dark matter particles can interact between themselves with a probability given by a certain cross section. This process tends to thermalize the inner halo, where the densities are higher and the collisions are therefore more frequent, and can lead to the natural creation of DM cores, which are possibly in a better agreement with the observations analyzed in the current work. Using the analytical expressions derived in \cite{yang24}, \citetalias{mancera24} provided an SIDM fit to the rotation curve of $\rm{AGC\ 114905}$, finding halo parameters that are as extreme as for the CDM model, but with the advantage that the low inner DM densities could be, in this scenario, naturally explained by the dark matter self interactions. \cite{nadler23} have indeed shown that SIDM simulations can more easily produce, with respect to CDM, halos with inner densities as low as those inferred from the observations of these UDGs. We are currently developing similar simulations to those presented in this paper, but including the dark matter self-interactions, with the plan of investigating whether these will lead to stable disc galaxies like $\rm{AGC\ 114905}$, naturally producing at the same time low-density halos consistent with the observations. These results will be presented in a future paper.

Regardless of the formation of these halos, note that our findings imply that even a disc with a baryon fraction higher than the cosmological average (our `Case 1') would not develop global instabilities. As discussed in depth in \citetalias{mancera24}, such a scenario could be due to different mechanisms related to a strong influence from the environment or to an atypical accretion history that would lead this galaxy to accrete more gas (previously expelled by nearby galaxies) with respect to the average. Given the isolation of $\rm{AGC\ 114905}$ and the absence of such large baryon fractions in current cosmological simulations, this scenario was deemed (and remains) unlikely. However, our stability analysis shows that such galaxies could exist in the Universe and therefore implies that these formation scenarios should not be ruled out and need to be investigated further.

\section{Summary and conclusions}\label{conclusions}
In this study, we performed idealised N-body simulations, using the code AREPO \citep[see][]{weinberger20}, of the HI-rich ultra diffuse galaxy $\rm{AGC\ 114905}$. We created a system composed of a stellar disc, a gas disc and a DM halo, initially in equilibrium with each other and with properties that resemble the recent observations of \citetalias{mancera24}. We mainly investigated four simulations, assuming two different DM halos of different masses (having very different baryon fractions but both leading to circular velocities consistent with the observations), with and without the inclusion of hydrodynamics. The scope of this work was to revisit the simulations from \citetalias{sellwood22} and to study whether a UDG with the physical properties as determined in \citetalias{mancera24}'s observations could or not be stable against global gravitational instabilities.

Our main findings are the following:
\begin{enumerate}
    \item for the four main simulations analyzed in this study, we found that both the gas and stellar discs of $\rm{AGC\ 114905}$ are stable against global instabilities and evolve unperturbed inside low-density DM halos. Even a model with a baryon fraction larger than the cosmological average leads to a stable galaxy;
    \item the apparent discrepancy with the results of \citetalias{sellwood22} is due to the fact that our gas and stellar discs are hotter and are embedded in slightly more massive DM halos, in agreement with the data from \citetalias{mancera24}. Models with a less significant contribution from the DM halo ($M_{\rm{DM}}(<10\ {\rm kpc})\ll 10^9\ M_{\odot}$) and with dynamically colder discs ($\sigma\approx5$ km s$^{-1}$) would lead to the development of global instabilities, as found by \citetalias{sellwood22}. Such models would however be in disagreement with the data from \citetalias{mancera24}; 
    \item intermediate models that include a DM halo with $M_{\rm{DM}}(<10\ {\rm kpc})\gtrsim 10^9\ M_{\odot}$, but have dynamically colder discs, are stable when hydrodynamics is included and develop instead global instabilities in the collisionless case. While this result must be confirmed by future simulations including also baryon physics, it points out how the proper modelling of these discs is crucial to understand their stability; 
    \item our results are generally in agreement with previously derived local and global instability criteria. We show however that, for galaxies close to instability, simple analytical criteria may be not accurate and numerical simulations are needed.

\end{enumerate}

The results of this paper show that $\rm{AGC\ 114905}$ (and most likely also other similar, less extreme UDGs, as we found for $\rm{AGC\ 242019}$) can survive and be stable inside a DM halo that observations suggest having very atypical properties in the context of the standard CDM framework \citep[see][]{mancera22,kong22,mancera24}. While the lack of stability has been evoked in the past as a reason for inconsistencies in the observational data (\citetalias{sellwood22}), we find that the stability of these UDGs is not an issue, removing doubts about the validity of the data and therefore confirming that such low-density DM halos, albeit very rare, can exist in the Universe. The next challenge will be to investigate how these peculiar halos (and galaxies) can form, by potentially exploring also alternative DM candidates, providing clues on the properties of dark matter itself.

\section*{Acknowledgements}
We are grateful to Haibo Yu, Daneng Yang and Cecilia Bacchini for stimulating conversations that led to some of the ideas discussed in this paper, and to the anonymous referee for their insightful comments. PEMP acknowledges the support from the Dutch Research Council (NWO) through the Veni grant VI.Veni.222.364. FM acknowledges funding from the European Union - NextGenerationEU under the HPC project `National Centre for HPC, Big Data and Quantum Computing` (PNRR-M4C2-I1.4-CN00000013-CUP J33C22001170001). This publication is part of the project `A modern Mercury anomaly as a testbed for dark matter' with file number OCENW.XS22.4.116 of the research programme Open Competition Domain Sience-XS, round 2022, which is financed by the Dutch Research Council (NWO).

\section*{Data Availability}
The optical and HI data used to create the initial conditions of the simulations are publicly available. The simulation data presented in this article are available upon reasonable request to the corresponding author. All the videos of the simulations run in this study are available at \href{https://doi.org/10.5281/zenodo.14852154}{this link}.




\bibliographystyle{mnras}
\bibliography{biblio} 




\appendix
\section{Dark matter evolution}\label{darkmatter}
\begin{figure}
   \includegraphics[clip, trim={0.1cm 0.1cm 0cm 0cm}, width=\linewidth]{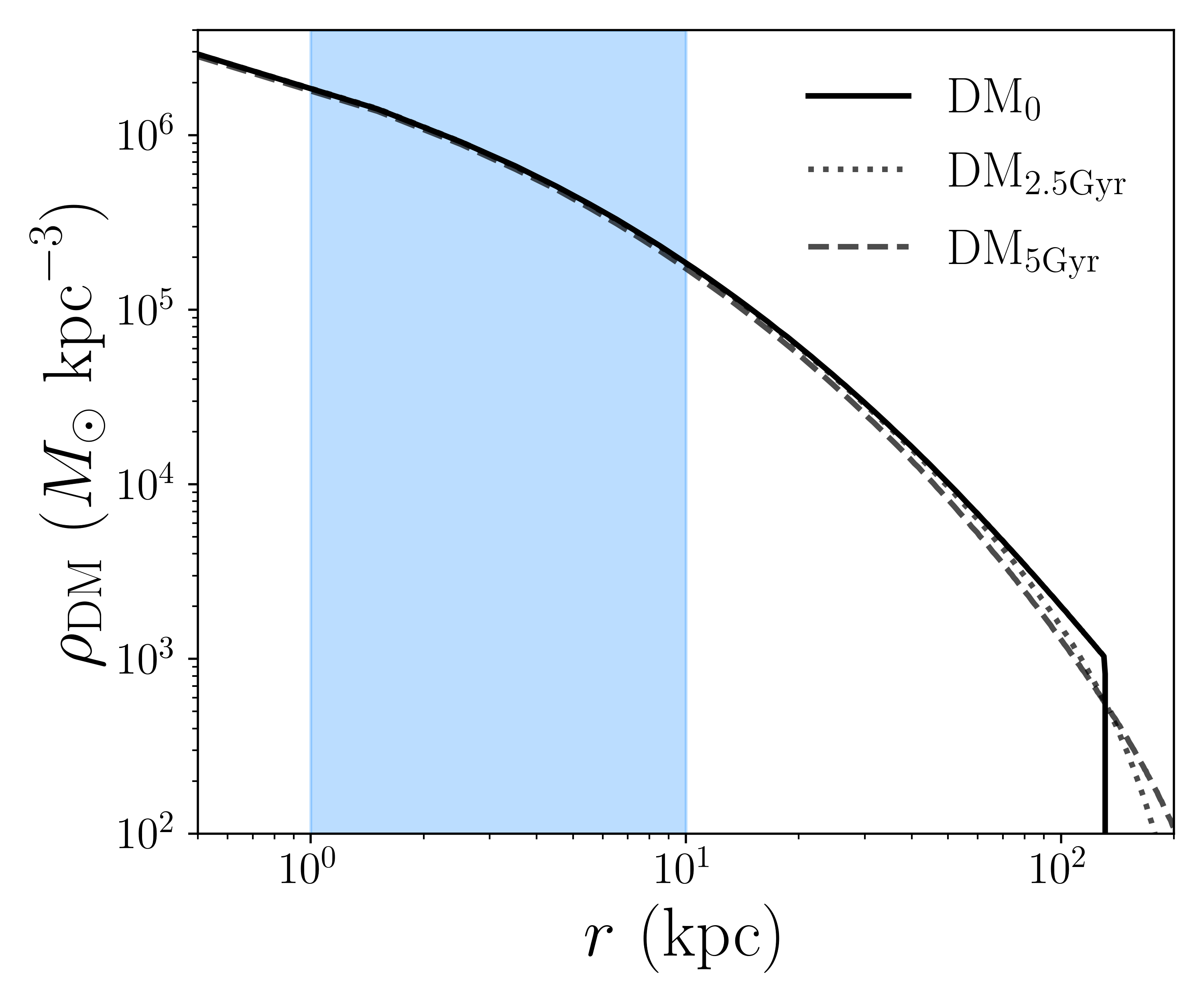} 
   \caption{Dark matter density profile extracted from the simulation `Case 2' with hydrodynamics, at the beginning (solid curve), at $2.5$ Gyr (dotted curve) and at the end of the simulation (dashed curve). The halo is initially truncated at $r=r_{\rm{t}}$ and as the simulation evolves the particles tend to redistribute themselves and to flow at $r>r_{\rm{t}}$. The blue vertical band marks the region where observational data are available.} \label{fig:dmprofiles}
    \end{figure}

In Figure~\ref{fig:dmprofiles} we show the evolution of the density profile of the DM halo extracted from the `Case 2' simulation with hydrodynamics, at the beginning, at $2.5$ Gyr and at the end of the simulation. The DM evolution in the other simulations is analogous to what shown here. We can see that at the start of the simulation the DM profile is truncated at a radius $r_{\rm{t}} = 2r_{\rm{vir}} = 136$ kpc, which is necessary to generate the initial DM particle realization, as explained in Section~\ref{numsims}. As the simulation evolves, the DM particles tend to naturally redistribute and to flow to larger radii, slightly lowering the densities below $r_{\rm{t}}$. At later times, some particles flow out of the computational domain (whose limit is defined in Figure~\ref{fig:dmprofiles} by the end of the plot $x-$axis, at $200$ kpc), lowering the total mass of the halo. This effect is however almost negligible, given that we chose a large enough simulation box. In the region of interest for this study (shown by the blue band, corresponding to the region where observational data are available), the changes in the DM profile are negligible and do not affect the circular (and hence rotational) velocity of the galaxy (see Figure~\ref{fig:kinematics}). Some minimal effects can be seen at about $20$ kpc from the center, where the change in DM density tends to reduce the gas rotational velocity by 1 or 2 km s$^{-1}$.\vspace{3cm}

\section{Components of the velocity dispersion}\label{dispComponents}
\begin{figure*}
   \includegraphics[clip, trim={0cm 0cm 0cm 0cm}, width=\linewidth]{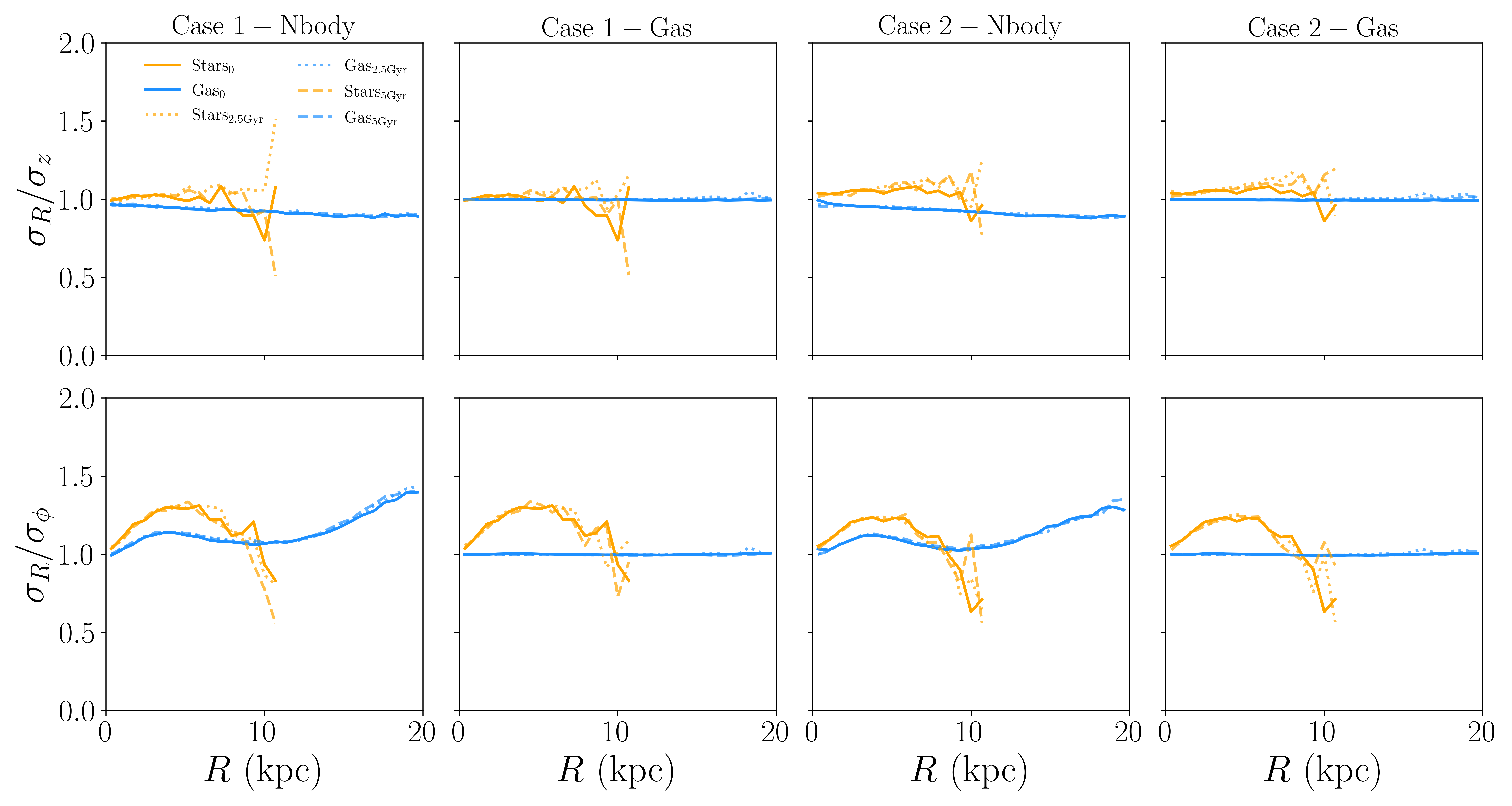}
   \caption{Ratios of the velocity dispersion of the stellar (orange) and the gas (blue) discs of the 4 main simulations analyzed in this study, at the beginning (solid curves), at $2.5$ Gyr (dotted curves) and at the end (dashed curves) of the simulation. The top panels show the ratios between the radial and the vertical components, while the bottom panels show the ratios between the radial and the azimuthal components. For the two cases with hydrodynamics (second and fourth column) we show, for the gas disc, the ratios of the three components of the total (thermal plus turbulent) velocity dispersion. In all simulations, the velocity dispersion of the stellar and the gas disc is nearly isotropic at all times.}
            \label{fig:Dispersions}%
   \end{figure*}

In Figure~\ref{fig:Dispersions} we show the ratios between the three components of the stellar (orange) and gas (blue) velocity dispersion in the four main simulations analyzed in this study (see Section~\ref{results} and Figures~\ref{fig:snapshots}--\ref{fig:kinematics}). As discussed in Section~\ref{SCM}, with AGAMA we built models in which both the stellar and the gas discs have velocity dispersion profiles that are roughly isotropic. This is shown by the solid lines in Figure~\ref{fig:Dispersions}, which represent the initial ratios of the velocity dispersion along the radial and vertical components (top panels) and along the radial and azimuthal components (bottom panels), for the `Case 1' (first two columns) and `Case 2' models (last two columns). Both discs exhibit velocity dispersion profiles that are nearly identical along the radial and vertical components. The azimuthal velocity dispersion has instead a slightly different profile with respect to the other two components, but the value of $\sigma_{\phi}$ is within $\sim20\%$ that of $\sigma_{R}$ and $\sigma_{z}$ everywhere in both discs in all simulations (except for $R\gtrsim15$ kpc in the gas disc of the two collisionless simulations, where the densities are very low and there are very few particles). Note that for the simulations with hydrodynamics (second and fourth columns) the gas velocity dispersion is dominated by the thermal velocity dispersion, which is by construction isotropic, and therefore the ratios are almost exactly equal to 1 everywhere.\\

Finally, Figure~\ref{fig:Dispersions} also shows that there is almost no evolution in the velocity dispersion ratios as the simulations evolve. This demonstrates how the velocity dispersion of both stars and gas (see also Figure~\ref{fig:kinematics}) remains isotropic for the entire simulation time, in line with the absence of strong gravitational instabilities (see Section~\ref{results}).
      
\bsp	
\label{lastpage}
\end{document}